\documentclass[usenatbib]{mn2e}
\usepackage{epsfig}
\usepackage{natbib}
\newcommand{\tvir}{T_{\rm vir}}

\newcommand{\tmax}{T_{\rm max}} 
 
\newcommand{\hmpc}{h^{-1}\,{\rm Mpc}}
\newcommand{\hpc}{h^{-1}\,{\rm pc}}

\newcommand{\msun}{{\rm M}_{\odot}}

\newcommand{\hubunits}{{\rm km}\;{\rm s}^{-1}\;{\rm Mpc}^{-1}}
\newcommand{\hkpc}{h^{-1}\;{\rm kpc}}

\newcommand{\mgal}{M_{\rm gal}}

\newcommand{\mhalo}{M_{\rm halo}}

\newcommand{\rvir}{R_{\rm vir}}

\newcommand{\newgad}{{\small GADGET}-2}

\def\msunyr{\msun\,{\rm yr}^{-1}}

\begin{document}
\title{Galaxies in a simulated $\Lambda$CDM Universe I: cold mode and
hot cores } 
\author[D. Kere\v{s} et al]
{Du\v{s}an Kere\v{s}$^{1, 2}$, Neal Katz$^2$, Mark Fardal$^2$, Romeel
  Dav\'e$^3$, David H. Weinberg$^4$\\
\\
$^1$Institute for Theory and Computation, Harvard-Smithsonian Center
  for Astrophysics, Cambridge, MA 02138,
 dkeres@cfa.harvard.edu\\
$^2$Astronomy Department, University of Massachusetts at Amherst, MA 01003; 
nsk@kaka.astro.umass.edu, fardal@astro.umass.edu \\
$^3$University of Arizona, Steward Observatory, Tucson, AZ 85721;
rad@astro.as.arizona.edu \\
$^4$Ohio State University, Department of Astronomy, Columbus, OH 43210;
dhw@astronomy.ohio-state.edu \\
}

\maketitle
\begin{abstract}

We study the formation of galaxies in a large volume ($50\hmpc$, $2 \times
288^3$ particles) cosmological simulation, evolved using the entropy
and energy conserving smoothed particle hydrodynamics (SPH) code \newgad.
Most of the baryonic mass in galaxies of all masses is originally acquired
through filamentary ``cold mode'' accretion of gas that was never shock
heated to its halo virial temperature, confirming the key feature of our
earlier results obtained with a different SPH code \citep{keres05}.
Atmospheres of hot, virialized gas develop in halos above $2-3\times 10^{11}
\msun$, a transition mass that is nearly constant from $z=3$ to $z=0$.
Cold accretion persists in halos above the transition mass, especially at
$z \geq 2$.  It dominates the growth of galaxies in low mass halos at all
times, and it is the main driver of the cosmic star formation history.
Our results suggest that the
cooling of shock-heated, virialized gas, which has been the
focus of many analytic models of galaxy growth spanning more than
three decades, might be a relatively minor element of
galaxy formation.
At high redshifts, satellite galaxies have gas accretion rates 
similar to central galaxies of the same baryonic mass, but at $z<1$ the 
accretion rates of low mass satellites are well below those of comparable 
central galaxies.  
Relative to our earlier simulations, the \newgad simulations predict much
lower rates of ``hot mode'' accretion from the virialized gas component.
Hot accretion rates compete with cold accretion rates near the transition
mass, but only at $z \leq 1$.  Hot accretion is inefficient in halos above
$\sim 5\times 10^{12}\msun$, with typical rates lower than $1\msunyr$ at
$z \leq 1$, even though our simulation does not include AGN heating or other 
forms of ``preventive'' feedback.  Instead, the accretion rates are low 
because the inner density profiles of hot gas in these halos are shallow, 
with long associated cooling times.  The cooling recipes typically used in 
semi-analytic models can overestimate the accretion rates in these halos by 
orders of magnitude, so these models may overemphasize the role of preventive 
feedback in producing observed galaxy masses and colors.  A fraction of the 
massive halos develop cuspy profiles and significant cooling rates between 
$z=1$ and $z=0$, a redshift trend similar to the observed trend in the 
frequency of cooling flow clusters.  

\end{abstract}


\section{Introduction}

The wealth of high quality observational galaxy data at
different epochs, from the earliest times to the present, has rapidly grown
in the last several years. This provides a great opportunity 
to calibrate and improve our theoretical understanding of galaxy formation and
evolution.
One of the remaining challenges in theoretical extragalactic
astronomy is the origin of the bimodality in galactic
properties: massive galaxies that are typically red, elliptical systems with
very little star formation, and
lower mass galaxies that are typically blue, disky, star forming systems
(e.g. \citealt{kauffmann03a, baldry04}). 
Usually one invokes various feedback processes to shut down star
formation in massive galaxies, which enables them to populate the red
sequence \citep[see review in][]{hopkins08b}. In addition,
without feedback in the low mass galaxies, a large fraction of
baryons would cool and form stars, even at very early times \citep{white78},
contrary to the small fraction of baryons in observed galaxies
\citep[e.g.][]{bell03}.

To understand where, when, and to what degree we need feedback during
galaxy formation and evolution, we first have to understand how galaxies are
supplied with baryons, and in particular with the gas that provides the fuel
for star formation.
In our previous work \citep[K05 hereafter]{keres05},
we showed that it is the smooth accretion of
intergalactic gas that dominates the global galactic gas supply, not accretion
by merging, and that it proceeds via two stages.
First, gas is accreted
through filamentary streams, where it remains relatively cold
before it reaches the galaxy \cite[K05]{katz03}. This process,
which we call cold mode accretion,
is very efficient, and because the gas does not need to cool it
falls in on approximately a free-fall time. This means that the baryonic 
growth closely tracks the growth of the dark matter halo, albeit with a slight
time delay. 
Cold mode accretion dominates the global growth of galaxies at
high redshifts and the growth of lower mass objects at late times. As the
dark matter halo grows larger, a larger fraction of the infalling material
shock heats to temperatures close to the virial temperature. In the denser,
central regions a fraction of this hot gas is able to cool.  
This process, virial shock heating followed by radiative cooling,
is the one highlighted in classic papers on galaxy formation theory
\citep[e.g.][]{rees77, silk77, white78, white91}, and
to differentiate it from the previous mode we call it hot mode accretion.
In massive halos, this hot mode accretion not only supplies massive galaxies
with fresh gas but could also be identified as cluster cooling flows
\citep[e.g.][]{fabian94}, although observations suggest that cooling in most
clusters does not actually reach down to
galactic temperatures, perhaps because of heating by AGN or some other
feedback mechanism \citep[e.g.][]{mcnamara07} .
The dominant accretion mode depends on halo mass, and the
transition between these two regimes occurs at a mass of about 
$M_{\rm halo}=2-3\times 10^{11}\msun$. At higher redshifts, cold filaments 
are able to survive within halos above this transition mass,
i.e. halos dominated by hot halo gas. Overall, the bulk of the 
baryonic mass in galaxies is accreted through the cold accretion mode.

Cold accretion owes its existence to the short cooling times present
in low mass halos near the virial radius, which prevents the
development of a stable  accretion shock.  The detailed criteria
required to prevent the formation of a virial shock in low mass
idealized systems have been derived by \citet{binney77} 
and \citet{birnboim03}. However, in cosmological simulations the
situation is more complex than in these simple models. For example, 
the spherical symmetry assumed in the case of \cite{birnboim03} will
not be valid owing to the presence of dense, cold filaments 
that dramatically enhance the density contrasts in the gas and
shorten the cooling times. 
In halos with masses near the transition mass from cold to
hot mode, this allows 
cold filamentary flows to still supply the central galaxy with gas
even though some of the infalling gas shock heats to near the virial 
temperature.

There are two main classes of galaxy formation feedback processes: those that
prevent
gas from entering a galaxy in the first place, ``preventive'' feedback,
and those that expel a fraction of
the gas that does manage to enter the galaxy, ``ejective'' feedback.
AGN ``radio mode'' heating is an example of preventive feedback
\citep[e.g.][]{best05,croton06} and winds driven by supernovae are
an example of ejective feedback \citep[e.g.][]{dekel86,springel03a}. 
The effectiveness of these two feedback types will vary depending on the
dominant accretion mode in the halo hosting the galaxy.
Because cold mode halos have very little halo gas outside of the cold dense
filaments, it is not clear if feedback processes can drastically affect
the accretion of gas. Ejective feedback, such as galactic winds driven by
supernovae, could still lower the masses of galaxies by expelling
already accreted material. 
Such winds, however, might be stopped by the quasi-spherical, hot halos
that surround hot mode galaxies. 
Conversely, preventive feedback, like the AGN radio mode,
is likely only effective in hot mode halos, where it can
prevent the hot, dilute gas that is in quasi-static
equilibrium from cooling. 
Similar constraints apply for the ``quasar wind mode'' of AGN
feedback \citep{dimatteo05}, which could possibly be a mixture of
these two feedback modes. At first, 
a relatively small fraction of galaxy mass is
ejected in the quasar wind, but later the energy released from the black
hole accretion could in principle provide ``preventive'' feedback if 
the quasar was active within a halo that has a hot quasi-spherical atmosphere.
In summary: ejective feedback is likely to be most
effective in cold mode halos, and preventive feedback is likely to be
effective only in hot mode halos.  
An exception to this rule is ``pre-heating'' of intergalactic gas
by photoionization \citep{efstathiou92, quinn96, thoul96},
or gravitational shocks
\citep{mo05}, which is a form of preventive feedback that will
mostly affect low mass, cold mode halos.

Current cosmological simulations have insufficient resolution
(often by large margins) to predict these feedback processes
from first principles, though they may incorporate simplified
recipes designed to achieve these effects.
It is, therefore, important to gain a general 
understanding of when and where these 
two classes of feedback processes are needed to explain the observed
properties of galaxies. 
In this paper, we examine the growth of galaxies by gas accretion
and the dependence of this process on galaxy mass, halo mass,
and redshift.
In a companion paper \citep[Paper II, hereafter]{keres09},
we compare the observable properties of our simulated galaxies, 
such as stellar masses
and specific star formation rates, to current data.

In K05 we showed that the mass of the transition from cold to hot mode
accretion  does not depend much on the resolution of the cosmological
simulation.  However, we showed that increasing
the mass and spatial resolution substantially reduced the amount of
hot mode accretion.  This suggested that the simulations from K05
had too much hot mode accretion, perhaps the result of numerical problems
inherent in the Smoothed Particle Hydrodynamics (SPH)
formulation that we used (see \citeauthor{springel02} \citeyear{springel02},
SH02 hereafter, for a brief review). 
How much gas is able to
cool from the hot atmosphere remained, therefore, an unanswered
question in K05. In this paper, we use the \newgad SPH code
\citep{springel05a} because it does not suffer from numerical
overcooling in massive halos, as we demonstrate in Kere\v{s} et al. (2009b,
in preparation) and in \citet{mythesis}, and because it parallelizes more 
efficiently to large numbers of processors.
The SPH algorithm implemented in \newgad, which
simultaneously conserves entropy and energy, is a reliable method for
simulations of cosmologically relevant volumes where galaxies are often only
resolved with a small number of particles.  In our previous simulations, this
low resolution allowed a mixing of the cold and hot gas
phases, which led to an artificial increase in the cooling rates and an
overestimate of the hot mode accretion \citep{pearce01, croft01}.
In our new \newgad simulations, we show that a large
fraction of the massive, hot mode halos cannot cool their gas, 
owing to nearly uniform density cores that develop in these
halos.  However, as discussed below (\S\ref{sec:acc_rates}), \newgad
simulations may still suffer numerical problems that distort
gas accretion rates, albeit at lower magnitude; fully understanding
these effects will require tests beyond those in this paper.
Relative to K05,
we also increase the simulation volume by about an order of magnitude,
allowing much better statistics for high mass halos and galaxies.
We use a similar but slightly better mass resolution. 

Here we present theoretical predictions of galaxy masses and accretion
rates in cosmological simulations without strong feedback. 
The energy from supernovae pressurizes the
interstellar medium, but it does not drive large scale outflows.
We do not include
any other feedback mechanisms that could prevent the accretion of gas or 
that could expel gas from galaxies in the mass range of interest in this paper.
The only exception is that we include an extragalactic UV background, which
affects the intergalactic medium and reduces gas accretion in the lowest
mass halos. 

In the Paper II we confirm earlier findings that the amount of
material that cools onto galactic component is too high in simulations
such as ours, which causes galaxies at any mass, but especially at the
low and high mass end to be more massive than their observed
counterparts. There we argue that an efficient high redshift feedback
in low mass galaxies (perhaps with the addition of low redshift AGN
feedback) is the key for getting the correct galaxy masses.

Many of the results in this paper and its companions (Paper II, and
Keres et al. 2009b (in preparation))
first appeared as part of D. Kere\v{s}'s PhD thesis at University of
Massachusetts, Amherst \citep{mythesis}.

In \S\ref{sec:simulations} we describe the basic parameters and setup of
our new simulations. In \S\ref{sec:accretion} we describe the smooth
accretion of gas, both globally and for individual objects. In
\S\ref{sec:halos} we describe the properties of non-cooling and
cooling halos and conclude in \S\ref{sec:conclusions}.  

\section{Simulations}
\label{sec:simulations}

We adopt a cold dark matter model dominated by a cosmological constant,
$\Lambda$CDM. In K05 we studied the properties of gas accretion in galaxies
and their growth using an older version of the $\Lambda$CDM
cosmology. 
Data from the first 3 years of the WMAP mission \citep{spergel07}
suggest much lower values of $\sigma_8$ and $\Omega_m$ 
(as anticipated by the galaxy clustering analyses of
\citet{vandenbosch03} and \citet{tinker05}),
which slightly changes
the time and redshift dependence of structure and galaxy growth.
We therefore use the following cosmological parameters: $\Omega_m=0.26$,
$\Omega_{\Lambda}=0.74$, $h\equiv H_0/(100\;\hubunits)=0.71$, and a
primordial power spectrum index of $n=1.0$. For the amplitude of the mass
fluctuations we use $\sigma_8=0.75$, and for the baryonic density we adopt
$\Omega_b=0.044$.
All of these cosmological parameters are consistent with recent
measurements from the WMAP team \citep{spergel07} and with various large scale
structure measurements\footnote{\tt 
  http://lambda.gsfc.nasa.gov/product/map/dr2/parameters.cfm (see the
  $\Lambda$CDM/All values)}, except the primordial power spectrum
index which is slightly higher in our simulations.
We model a 50.0$\hmpc$ comoving periodic cube using $288^3$ dark
matter and $288^3$ gas particles, i.e. around 50 million particles in total.
Gravitational forces are softened using a cubic spline kernel of
comoving radius $10\hkpc$, approximately equivalent to a Plummer
force softening of $\epsilon_{\rm grav} = 7.2\hkpc$. 
Using the naming scheme from K05, we will refer to this
simulation as L50/288 later throughout the text. We occasionally compare
this simulation to a \newgad simulation of a smaller volume simulation,
L22/128. This simulation uses the same cosmological parameters and
initial conditions as those that were used in K05. To test for resolution 
effects we also use L11/64 and L11/128 simulations,
again with the same initial
conditions as their analogs in K05 but simulated with \newgad. The
parameters of all of these simulations are shown in
Table~\ref{tbl:sims}.  For the visualizations of cold accretion
in \S\ref{sec:illustrations} below, we use a higher resolution
simulation, L12.5/288, which is described there.

\begin{table*}
\begin{tabular}{ccccccc}

\hline Name&$L (\hmpc)$&$N$&$z_{\rm fin}$&
$m_{\rm gas}$($\msun$)&\\ 
\hline 
\bf{L50/288}&$50$&$2\times 288^3$&$0$&
$9\times 10^{7}$&\\ 
L22/128&$22.22$&$2\times
128^3$&$0$&$1.1\times 10^8$&\\ 
L11/64&$11.11$&$2\times
64^3$&$1$&$1.1\times 10^8$&\\ 
L11/128&$11.11$&$2\times
128^3$&$1$&$1.3\times 10^7$&\\ 
\hline
\end{tabular}
\caption{Details of the simulations used in this paper. $L$
is the comoving box size, $N$ is the total number of particles
(dark+baryonic), $z_{\rm fin}$ is the final redshift to which the
simulation has been evolved and $M_{\rm gas}$ is the initial mass of
gas particles in these simulations
}
\label{tbl:sims}
\end{table*}

The initial conditions are evolved using the SPH code \newgad
\citep{springel05a}. The calculation of the gravitational force
is a combination of the Particle Mesh algorithm
\citep[e.g.][]{hockney81} for large distances
and the hierarchical tree algorithm \citep{barnes86, hernquist87} for
small distances. 
The smoothed particle hydrodynamics algorithm \citep{lucy77, gingold77}
used here is entropy and energy conserving, and it is based on the
version used in \citet{springel02}. We use a modified public version of the 
\newgad. Modifications are made to include the cooling, the uniform
UV background and the two-phase star formation algorithm. We briefly
describe these below. 

We include all the relevant cooling processes with primordial abundances as
in \citet{katz96}. 
We do not include any metal enrichment or cooling processes associated
with heavy elements or molecular hydrogen.
In all of these simulations we include a spatially uniform,
extragalactic UV background that heats and 
ionizes the gas. The redshift distribution and spectrum of this background is
slightly different than in K05. The background flux starts at $z=9$ and is
based on \citet{haardt01}. For more details about the calculation of
this UV background see \citet{oppenheimer06}. We note, however, that
smaller volume simulations of comparable resolution with our new
UV background and with the version used in K05 showed no noticeable
differences in the evolution of the galaxy population above our
resolution limit in the redshift range of interest in this paper,
$z=0$ to 4.

Once a gas particle reaches a density above the star forming threshold,
star formation proceeds in a sub-resolution two-phase medium where 
supernova energy
released by type II SNe during star formation balances
cold cloud formation and cloud evaporation by the hot medium as in
\citet{mckee77}.  Pressurization of the interstellar medium by SN feedback
enables more stable gas rich disks, but it does
not produce a large scale outflow, i.e. a galactic wind. 
We use the same star formation model
parameters as in \citet{springel03a}, which were
selected to match the $z=0$ relation between star formation
rate and gas surface density
(\citealt{kennicutt98}; see also \citealt{schmidt59}).  The
code calculates the threshold density for star formation as the density
where the mass weighted temperature of the two phase medium equals
10,000K. In practice this threshold density is constant in physical
units during the simulation,
and it corresponds to a hydrogen number density of
$n_h=0.13 {\rm cm}^{-3}$.
Each gas particle within the two-phase medium has an assigned star
formation rate, but the actual conversion from gaseous to stellar
particles proceeds stochastically \citep{springel03a}. This is similar to the
algorithm in \citet{katz92a}, where each
star particle takes half of the initial mass of a gas particle.

To identify bound groups of cold, dense baryonic
particles and stars we use the program SKID\footnote{
\tt http://www-hpcc.astro.washington.edu/tools/skid.html} 
(see K05 for more details).
Briefly, a galaxy identified by SKID contains bound stars and gas with an
overdensity $\rho/\bar{\rho}_{\rm baryon} > 1000$
and temperature $T < 30,000\,$K. Here, we slightly modify
this criterion and apply a higher temperature threshold at
densities where the two-phase medium develops.  Such a modification
is necessary to allow star forming two-phase medium particles
to be part of a SKID group,  since at high densities the mass-weighted
temperature in the two phase medium can be much higher than 30,000 K.
To identify halos we use both the Friends of Friends (FOF) and
Spherical Overdensity (SO) algorithms with the same parameters as in
K05, adjusted to the new cosmology. 

\section{Accretion rates}
\label{sec:accretion}

The baryonic masses $M_{\rm gal}$ of SKID galaxies include
stars, star-forming (two-phase medium) gas, and gas with
$\rho/\bar{\rho}_{\rm baryon} > 1000$ and temperature 
$T < 30,000\,$K. 
Using the convergence tests in \citet{murali02} as a guide, we adopt 
$M_{\rm gal}=64 m_{\rm SPH}$ as our nominal resolution limit,
above which galaxy baryonic masses are reasonably well converged.
For the L50/288 simulation, $m_{\rm SPH}=9.1\times 10^7 \msun$,
making the resolution criterion $M_{\rm gal} \ge 5.8\times 10^9 \msun$.
To define accretion rates we require a galaxy to be resolved
at both ends of the time interval used to calculate the rate.
This leaves us with around $10,000$ resolved galaxies at $z \le 2$.
To avoid counting the accretion of sub-resolution groups as a smooth
accretion, we define smooth gas accretion as the accretion of gas
particles that were not part of any SKID identified group at the previous
time, i.e. even bound groups smaller than 64 particles will not be
counted as smooth gas accretion. In practice this procedure will
avoid counting sub-resolution mergers down to galaxies with $\sim 20-30$
particles, i.e. $\sim 2\times 10^{9} \msun$, approximately the value
where the galaxy mass function in our L50/288 simulation
drops rapidly owing to our limited resolution and the UV background.

As in K05, we follow the temperature history of the accreted particles,
noting the maximum temperature a gas particle reaches before it becomes 
part of a galaxy, $\tmax$. We ignore the temperature while a particle
is in the two-phase medium for gas particles that  are ``recycled''
from one galaxy to another.
Following K05, we also split the smooth accretion into two components,
that with $\tmax < 250,000$K, representing cold mode gas accretion, and
that with higher $\tmax$, representing hot mode accretion. 
K05 found this to be a good empirical division between the
cold and hot accretion modes, and they argued that a cut in 
physical temperature describes the simulations better than a
cut in scaled temperature $\tmax/\tvir$.

\subsection{Global accretion rates}

\begin{figure*}
\epsfxsize=5.25in
\epsfbox{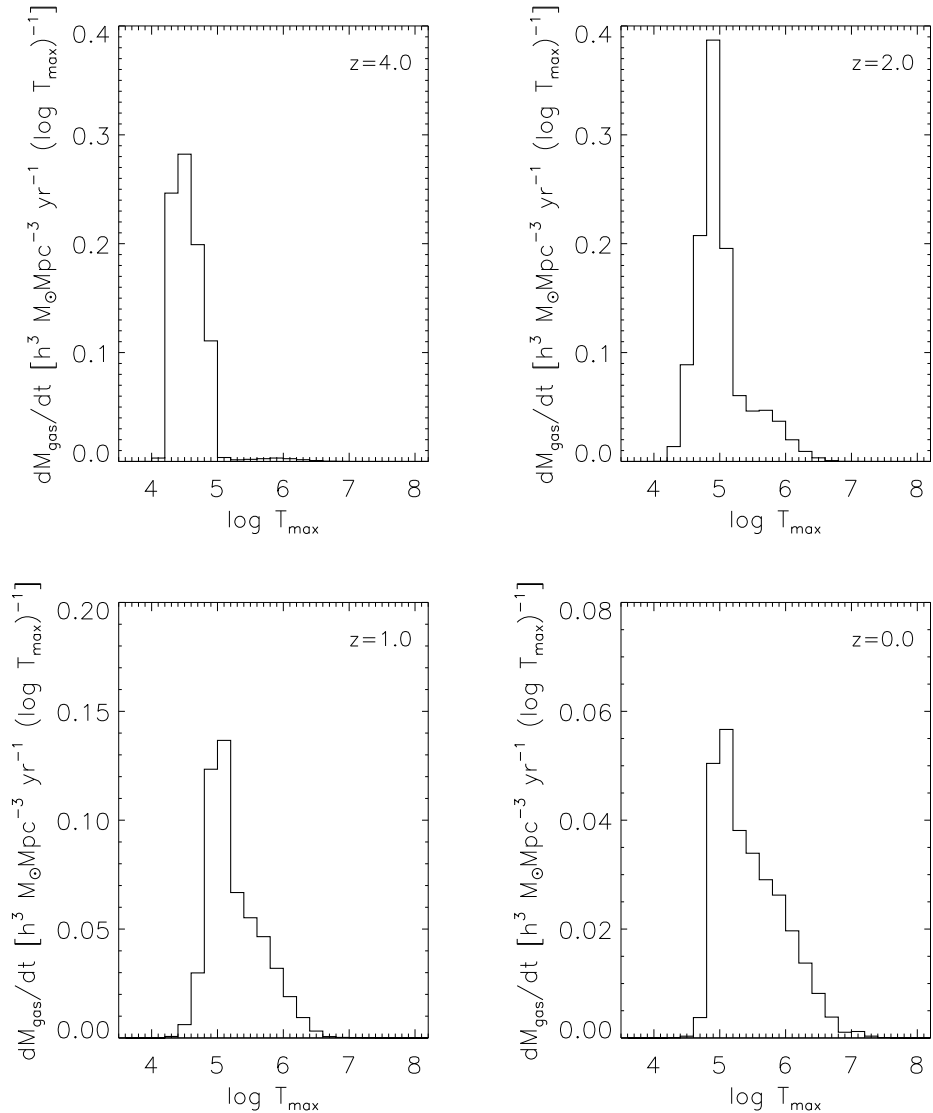}
\caption[The global volume averaged smooth gas accretion rates as a function
  of $\tmax$ for the L50/288 simulation.]{The global volume averaged
  smooth gas accretion rates as a function   of $\tmax$ for the
  L50/288 simulation. The histograms are calculated by adding the mass of
  all gas particles that were smoothly accreted in a given $\tmax$ bin
  between two simulation outputs. Note that we express the
  accretion rates per decade in  $\tmax$, unlike in K05 where they were
  expressed per bin. 
  }
\label{fig:tmax}
\end{figure*}

We calculate the global smooth gas accretion rates as the sum of gas
accreted by all resolved galaxies at a given time divided by the simulation
volume.
Following K05 we show results for the $\tmax$ distribution of the
global accretion in Figure~\ref{fig:tmax}. 
Global accretion is dominated by the cold mode accretion peak,
i.e. by gas accreted with $\tmax \le 250,000 K$, at all
redshifts including  $z=0$. 
The temperature minimum that separates cold from hot mode accretion
washes out after $z\sim2$, caused by the very weak hot mode
accretion in this new simulation, as we discuss below. As mentioned earlier,
SPH codes without second-order entropy conservation
can greatly overestimate hot mode accretion (SH02). 
This problem was hinted at in the high resolution simulations evolved with
P-TreeSPH in K05. A careful comparison between simulations 
that use the same cosmological parameters and initial
conditions but different SPH formulations shows that the
much lower hot mode accretion rates we find in
this new simulation are mostly the consequence of the different SPH
formulation (Kere\v{s} et al. 2009b, in preparation). 
We see that the global accretion rate declines towards lower redshift, as
we also saw in K05.
The change in cosmology also changes the magnitude of the global
accretion rates, making high redshift accretion less pronounced than in
K05 because low $\sigma_8$ causes structure to form later. Later halo
formation, slightly higher mass resolution, and a higher 
baryon-to-dark-matter 
ratio all contribute to the higher cold mode accretion rates seen at low
redshift. 

\begin{figure*}
\epsfxsize=4.5in
\epsfbox{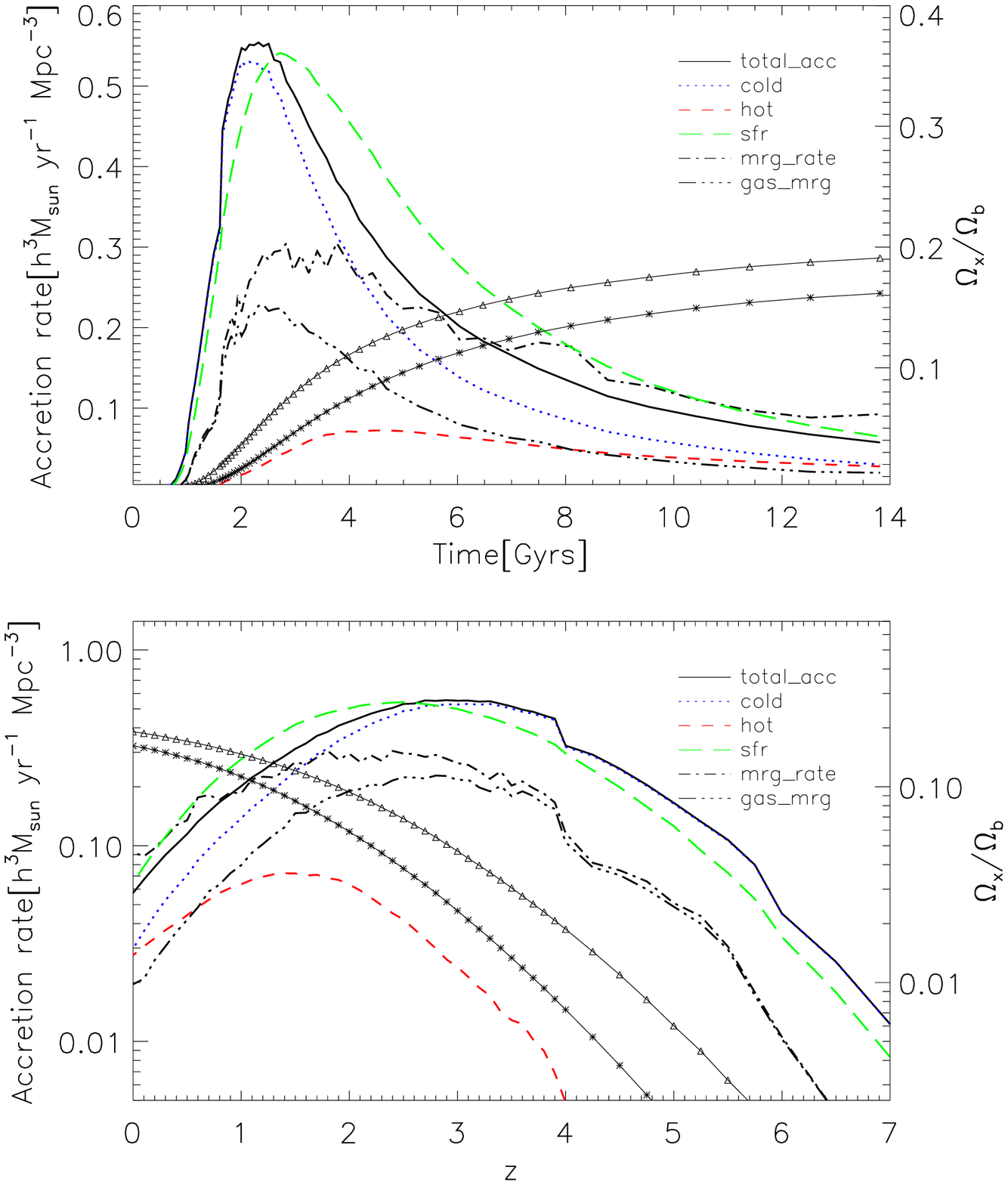}
\vskip -10pt
\caption[The volume averaged smooth gas accretion, star formation and
  merger rates of the resolved galaxy population in the L50/288
  simulation]{The volume averaged smooth gas accretion, star formation
  and merger rates of the resolved galaxy  population in the L50/288
  simulation. We show the volume averaged total smooth gas accretion
  rates (solid), smooth accretion in cold mode (dotted, blue) and smooth
  accretion in hot mode (dashed, red). We also plot the star formation
  rate densities 
  of the resolved galaxy population (green, long-dashed), the
  volume averaged rates of  mass increase through mergers (dot-dashed)
  and the gaseous part of the merger rates (tripple-dot-dashed). 
  The upper panel plots the rates versus cosmic time and the lower
  panel plots the rates versus redshift. In both panels we also overplot
  the fraction of baryons locked in resolved galaxies (triangles) and
  the fraction of baryons locked in stars in resolved galaxies (stars),
  with the corresponding values on the right hand side y-axis.
}
\label{fig:global_acc_j50}
\end{figure*}

The change in magnitude of the accretion rates with time is better
illustrated in  
Figure~\ref{fig:global_acc_j50}, where we show the total smooth gas accretion
rates, the cold mode accretion rates, and the hot mode accretion rates. 
The accretion rates peak around $z=3$ followed by a drop to $z=0$ of about a
factor of 10. The high 
redshift accretion rates at $z\sim 4-5$ are lower by about a factor
of 2-3 in L50/288 compared to K05.
The much lower accretion rates result from the later formation time of halos
at a given mass, which in turn results from the lower
value of $\sigma_8$ in the new WMAP cosmology. 
At high redshifts, this later halo formation time results in a smaller
number of galaxies that are included in the resolved galaxy sample. 
Since high redshift accretion is completely dominated by efficient
cold mode accretion, the magnitude of these rates approximately reflects
the growth of the dark matter halos that host the resolved galaxies (minus the
small contribution from sub-resolution mergers), as demonstrated in K05. 

The fast increase in the number of galaxies that cross the resolution limit
causes the sharp rise in global accretion from $z \sim 7-8$ to $z \sim 3$.
(Note, however, that we only count accretion onto pre-existing galaxies,
so the first appearance of a galaxy above the threshold does not
contribute to the accretion rate.)
As time increases, the typical accretion rate at fixed galaxy mass declines,
but at high redshifts this trend is overwhelmed by the larger number
of resolved galaxies and by the increase in average galaxy mass
(with higher mass galaxies having higher mean accretion rates).
At lower redshifts,
the number of resolved galaxies does not change significantly, and the
decrease in the average accretion rate at fixed galaxy mass
causes a drop in the global accretion rate.

In Figure~\ref{fig:global_acc_j50} we also compare the
rate of mass growth through mergers to that through smooth accretion. Here,
we include as mergers all the material that is added to
resolved galaxies between two simulation outputs that is not
smoothly accreted. Almost all of the material that is counted as
mergers joins the galaxies in dense baryonic clumps. 
Similar to the findings in K05, mergers dominate the
total mass growth at low redshift, in this case after around $z=1$. 
The gaseous part of the merger rates is now significantly higher than
in our previous work. This is largely caused by counting the
accretion of small, under-resolved, gas rich galaxies as mergers and
in part by the later formation of halos and galaxies, which keeps them
slightly more gas rich at late times. However,
the gas added through mergers is always significantly lower than that
added through smooth accretion. In a global sense, therefore, 
galaxies are supplied with gas, the fuel for star formation,
mainly by smooth accretion.

As in K05, the global star formation rate closely follows the smooth gas
accretion rate with a
relatively short time delay. This is a consequence of the 
continuous gas supply and the relatively short, observationally motivated
star formation time scale we adopt here. On a galaxy by galaxy
basis, there are systematic offsets from this tight relation between the
star formation rate and the smooth gas accretion rate, which we will
discuss later in this section (see \S\ref{sec:sfr_acc}).
Interestingly, the $z=0$ global SFR
is within a factor of two of the observed value \citep{hopkins06} even though
we do not include either supernova winds or AGN feedback.
At higher redshifts the observational estimates
have an extremely wide range, and
while our predicted star formation rates lie within with this range, 
the uncertainty of the comparisons prevents
strong constraints on the models.

However, the observed star formation history is quite uncertain 
  and, therefore, matching it at some given time is not necessarily a
  major success of a given model. A more constraining measure of
  previous baryonic mass accumulation is the amount of baryons that
  are locked in the galactic component. 
  We plot the mass accumulation history of baryons in resolved
  galaxies as a fraction of baryonic density of the universe in
  Figure~\ref{fig:global_acc_j50}. At $z=0$, in our L50/288 simulation
  around 20 percent of available baryons reside in galaxies and around 17
  percent of all baryons are locked in stellar component (in and out
  of galaxies). For the resolved galaxies only, these numbers are 19 and 16
  percents, respectively. This is factor of 3 higher than what is observed
  \citep{bell03}, which suggests that, globally, too much gas cools
  onto galaxies in our simulations. This problem is the main topic of
  Paper II, where we explore this discrepancy 
   more quantitatively and suggest what kind of feedback is needed to
  reconcile these differences. Here we only note that the amount of
  cooled baryons in the simulation analyzed here is lower that what was
  suggested in earlier work based on SPH simulations without ejective
  feedback \citep[e.g.][]{benson01}. A large part of this difference
  owes to the  use of the entropy and energy conserving
  code that  prevents  numerical overcooling in our paper, but
  additional effects likely owe to the lack of metal cooling in our
  simulation, the inclusion of a UV background which
  suppresses the formation of the smallest objects, and a lower $\sigma_8$
  which delays the formation times of halos.

\subsection{Virialization of the halo gas}
\label{sec:virialization}
\begin{figure*}
\epsfxsize=5.7in
\epsfbox{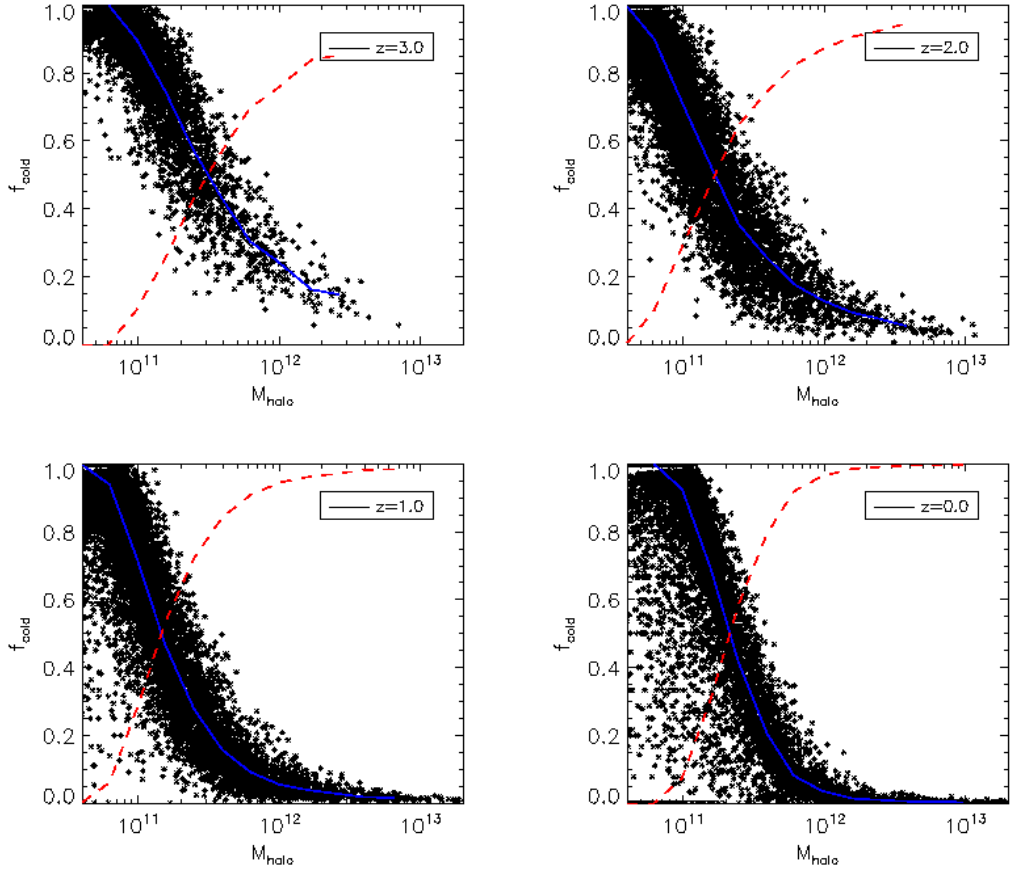}
\caption[]{The fraction of cold halo gas in a halo of a given
  mass at four different redshifts. Lines show the running median of
  the cold gas fraction (blue, solid) and the hot gas fraction
  (red, dashed). (We exclude any galactic gas.) 
}
\label{fig:cold_hfractions}
\end{figure*}

Before we proceed to discuss the accretion rates of individual
galaxies in our simulation, it is interesting to check 
if the transition mass between halos dominated by cold
filamentary flows and hot virialized gas occurs at a
mass similar to that in K05, i.e. $2-3 \times 10^{11}\msun$. 
For this purpose, Figure~\ref{fig:cold_hfractions} plots
the fraction of halo gas (excluding gas in SKID-identified galaxies) 
that is below 
the temperature that we use to divide cold and hot mode accretion, 250,000 K.
We see that cold halo gas remains the dominant component in
halos with masses lower than $2-3 \times 10^{11}\msun$, while for higher
mass halos hot gas begins to dominate, and the most massive halos are
almost completely filled with hot gas. The transition mass between cold and
hot halos is almost
constant with redshift even though the virial temperature of a halo of a
given mass increases with redshift roughly as $\tvir(z) \propto (1+z)$.
One can also see that hot, virialized
gas builds up gradually, with the transition occurring over more than an
order of magnitude in halo mass. There
is a large density contrast between filaments and the inter-filamentary
region, which allows cold mode accretion to coexist with hot gas
in the regions between the filaments in this transition mass range.
The spherical model of \citet{birnboim03} predicts an approximately
similar transition mass, smaller by a factor of $2-3$ (though this
depends on the precise definition of the transition in their model).
Even in this idealized model, the virial shock expands
outwards as the halo mass increases, approximately mimicking the
simulations. However, their model does not account for the
filamentary structure of the infalling gas, which makes the transition between
cold and hot halos sharper than in the cosmological simulations.
We also compared the halo cold gas fractions in the simulations
used in K05 with the L22/128 and L11/128 simulations, 
which used the same initial conditions
but were evolved with \newgad, and we find very similar results. 

In summary, as 
the halo mass increases the hot virialized atmosphere slowly
accumulates until it becomes the dominant halo gas component in massive
halos. In agreement with K05, 
we visually confirm (see \S\ref{sec:illustrations})
that cold filamentary streams survive in
halos with masses much larger than the transition mass, 
especially at high redshifts, which makes the transition from 
cold to hot domination a gradual one and makes the cold gas fraction
in massive halos slightly increase with redshift.
\cite{ocvirk08} and \cite{dekel09} show similar results
obtained with an adaptive refinement mesh hydrodynamic code.
At lower redshifts, the cold-to-hot transition is 
sharper, but it is also more sensitive to our precise definition because
the virial temperature of the halos at the transition mass approaches 
$250,000\,$K.
\citet{croton06} suggest that the transition masses in
K05 are biased high owing to the ``geometric'' SPH method in P-TreeSPH that
was used to evolve the simulations. 
The results from the energy and entropy conserving \newgad simulation that we
present here demonstrates clearly that this transition mass is not
sensitive to the SPH method used.
The transition mass depends, however, on the fraction of halo baryons
available in a halo and on the gas metallicity as discussed by
\citet{birnboim03}. In our L50/288 simulation, in halos
with $10^{11} \msun < M_h < 5\times 10^{11} \msun$, typically around 40 percent
of the halo baryons are residing out of the galactic component.
Our preliminary results from simulations that include the metal line cooling 
and supernova driven galactic winds, which increase the amount of gas in 
halos, indicate that the transition between halos dominated by 
cold and hot gas shifts upward by a factor $\sim3$ in halo mass.

\subsection{Illustrations of Cold Mode Accretion}
\label{sec:illustrations}

\begin{figure*}
\epsfysize=5.7in
\epsfbox{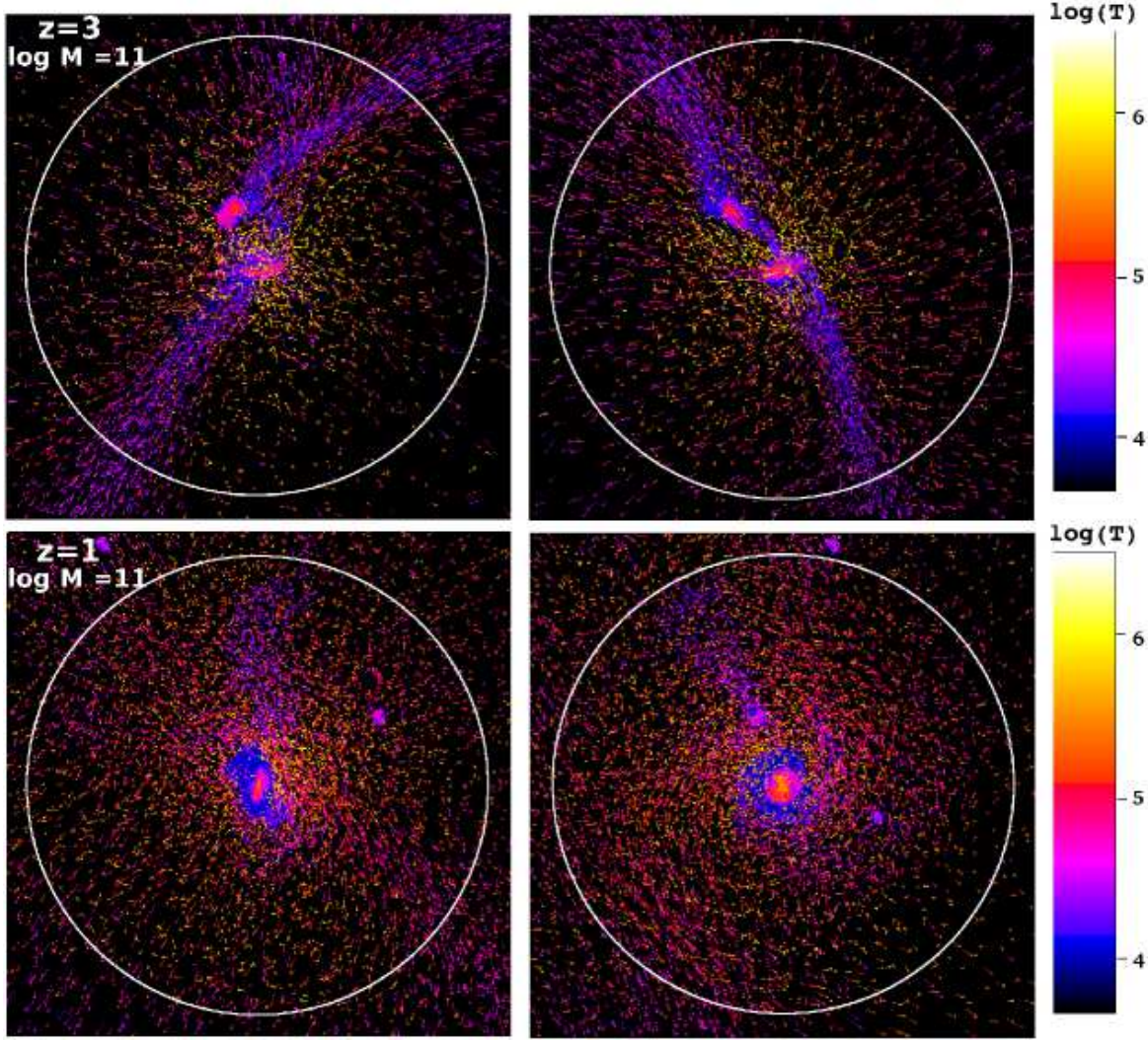}
\caption{Temperature of the gas in low mass halos at $z=3$ and $z=1$.
Panels show the gas in a region of $250 \hkpc$ (comoving) on a side
and $250 \hkpc$ (comoving) in projected depth. 
The virial radius is shown as the circles. Upper
panels show the two different projections of a halo at $z=3$ dominated by a
cold halo gas with $M_h=1.1\times 10^{11}\msun$, while lower panels
show a halo at $z=1$ with $M_h=1.12\times 10^{11}\msun$. Vectors attached
to particles show their projected velocities. Note that warmer gas temperature
within the densest clumps is caused by sub-resolution implementation
of the two-phase star forming medium. } 
\label{fig:lowmass}
\end{figure*}

\begin{figure*}
\epsfysize=8.7in
\epsfbox{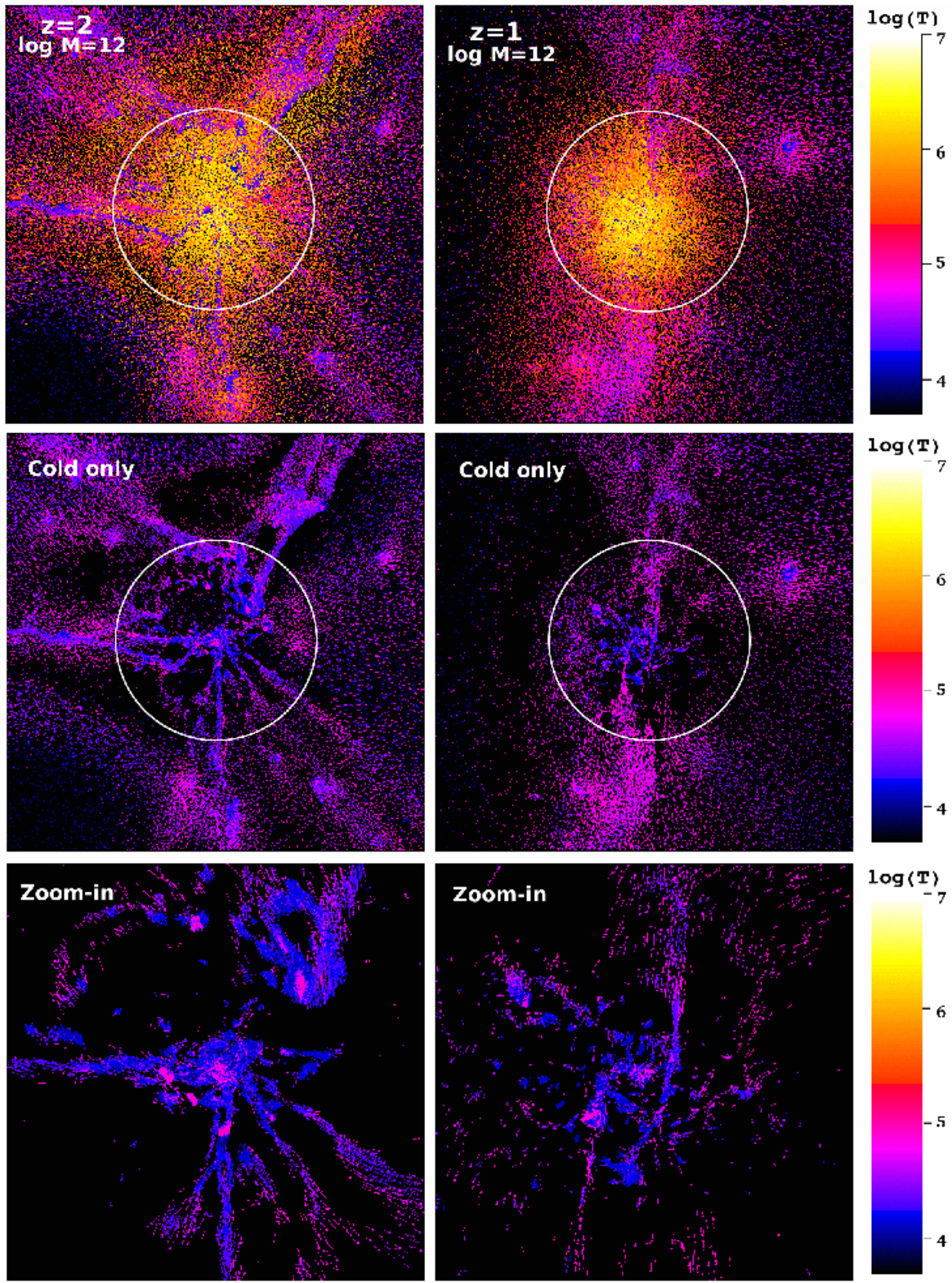}
\vskip -10pt
\caption{Temperature of the gas in Milky Way mass halos at $z=2$ and $z=1$.
Panels show the gas in a region of $1 \hmpc$ (comoving) on a side and
$1 \hmpc$ (comoving) in projected depth. The virial radius is shown as
the circles. 
Left panels show an $M_h=1.1\times 10^{12}\msun$ at $z=2$, while the
right panels show an $M_h=1.2\times 10^{12}\msun$ halo at $z=1$.
Upper panels show all gas particles. The middle panels show only the
gas with $T < 10^5 K$ but with the same color scale (indicated on the
right). The lower panels also show only the low temperature gas, but
are zoomed-in to show a region $375 \hkpc$
(comoving) on a side (approximately within $0.8 \rvir$) and $125 \hkpc$ 
(comoving) deep. 
Vectors attached to particles show their projected velocities. 
Note that warmer gas temperature
within the densest clumps in middle and lower panels is caused by 
sub-resolution implementation of the two-phase star forming medium. 
}
\label{fig:massive}
\end{figure*}

\begin{figure*}
\epsfxsize=3.6in
\epsfbox{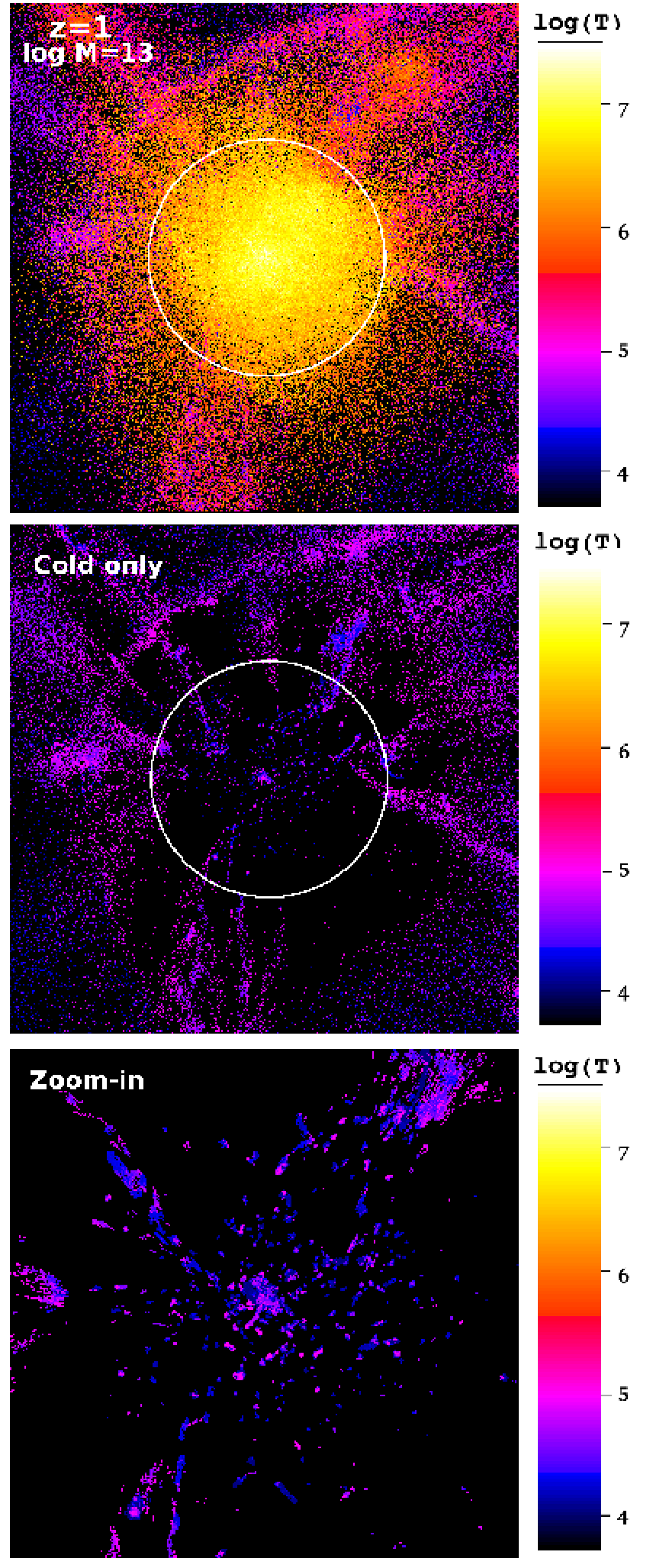}
\caption{
Temperature of the gas in a group size halo at $z=1$, 
with $M_h=9\times 10^{12}\msun$. 
Panels show the gas in a region $2 \hmpc$ (comoving) on a side and
$2 \hmpc$ (comoving) in projected depth. The virial radius is shown as
the circles. 
Upper panels show all gas particles. The middle panels show only the
gas with $T < 10^5 K$ but with the same color scale (indicated on the
right). The lower panels also show only the low temperature gas, but
they are zoomed-in to show a region $750 \hkpc$
(comoving) on a side (approximately within $0.8 \rvir$) and $250 \hkpc$ 
(comoving) deep.
Vectors attached to particles show their projected velocities.
  }
\label{fig:mostmassive}
\end{figure*}

How do filaments that are
driving the cold mode accretion connect to galaxies and halos?
How do these connections depend on halo mass and redshift?
The answers to these questions depend somewhat on numerical resolution.
In higher resolution simulations, the central cores
of filaments are able to reach higher densities and
therefore cool faster and become thinner. Furthermore, the low mass
galaxies and subhalos that can be tracked at higher
resolution can drag or disturb cold gas filaments in their parent halos.
To investigate the qualitative behavior of filamentary cold accretion,
we therefore use the highest resolution simulation currently
available to us, evolved to $z=1$ with $288^3$ particles
in a $12.5\hmpc$ comoving box (L12.5/288).
This simulation has no ejective feedback, 
but it was evolved with metal line cooling and
stellar mass loss that gets recycled into the surrounding gas. The
detailed properties of the galaxies that form in this simulation will
be presented  in future work, but it is worth mentioning that this
extra physics does not make a significant difference to the properties 
of the intergalactic halo gas, since without winds the metals and 
recycled material stay within or close to galaxies.
We explicitly checked whether the buildup of the hot gas in
halos changes with this additional physics and found no difference,
i.e. the transition mass between cold gas and hot gas dominated
halos is the same as  in all the other simulations used in this paper. The
gravitational softening  in this simulation is $220 \hpc$ (Plummer equivalent)
physical at $z=3$ and it increases to $440 \hpc$ physical at $z=1$. 
Such high resolution is necessary to see the variety of structures in
the cold accretion flows and to properly resolve the regions very close
to the galactic disks.

Figure~\ref{fig:lowmass} shows the characteristic structure of cold mode gas
in halos below the transition mass:
a halo at $z=3$ with $M_h=1.1\times 10^{11}\msun$ (upper panels) and 
a halo of the same mass at $z=1$ (lower panels). 
The particles are color coded by temperature, and their relative
density on the plot indicates the projected column density
(because the masses  of the gas particles out of galaxies are usually
identical).  
To illustrate the filamentary geometry, we show two different
projections of each halo.

As emphasized in K05, owing to the presence of the dense filamentary
streams, a typical high redshift halo has regions of great density
contrast.  High densities and  short cooling times in these
filaments make conditions unfavorable for shock propagation
\citep{binney77, birnboim03}.  In the inter-filamentary
regions, where the gas density is often one or two orders of magnitude
lower, some gas shock heats even when the halo mass is much lower
than the transition mass. This is in contrast with simple 1D models,
which necessarily average away density variation at a given radius,
and which typically show a quick transition from cold mode to hot mode.
In the upper panels of Figure~\ref{fig:lowmass},
dense filamentary flows connect to the outskirts of the central galaxy 
and to an infalling satellite galaxy (see \S~\ref{sec:satellites}). 
Tidal interaction has created a bridge between these two galaxies.
Near $\rvir$ the filaments are thicker than the galaxies themselves,
but they are then compressed by the surrounding shock heated (virial
temperature) gas, which extends to $\sim 0.5\rvir$.  Near the galaxies,
the filaments are narrower than the galactic disks.

The situation changes at low redshift, where lower mass galaxies 
typically reside in larger filamentary structures comparable in
cross section to the size of the halo. However, the densest parts of the
filaments can still survive within the virial radius. Deeper in the
halo, some of this gas gets shock heated to the virial temperature and some
stays cold, resulting in a mixture of cold and hot accretion that
supplies the gaseous galactic disk visible in the lower panels of
Figure~\ref{fig:lowmass}. 

Figure~\ref{fig:massive} shows more massive halos, with mass $M_h \approx
10^{12}\msun$, a factor $\sim 3$ above the transition mass seen in
Figure~\ref{fig:cold_hfractions}.
The 
limited volume of our L12.5/288 simulation does not contain any halo
of this mass at $z=3$ so we show halos at $z=2$ and $z=1$.
Here the situation changes dramatically. At $z=2$, a typical $\sim 10^{12}
\msun$ halo evolves at the intersection of several filamentary
structures. The shock heated gas in this halo fills up the cavities
between the filaments, and in some directions it extends to several virial
radii. One can clearly see how gas that falls onto a halo from the
region between the filaments gets heated at a shock
front, typically slightly outside $\rvir$.
The middle and the lower panels show only the gas with $T < 10^5 \rm K$, to
reveal the penetration and filamentary structure of the cold accretion.
These filaments are enhanced by infalling substructures,
which drags parts of the filaments into the more massive
halos and often splits larger filamentary structure into multiple
flows separated by hot gas.
Furthermore, halos of this mass are dominated by hot gas at
$T\sim\tvir$, which provides a high pressure environment that 
compresses the filaments. This results in the thin filamentary streams,
most of which can penetrate to the halo's central regions.
The zoom-in panel of the $z=2$ halo shows that these cold filaments
survive to supply both the central and the satellite galaxies 
(in the upper right corner of this panel) with cold
gas. However, a large fraction of the infalling gas is now shock
heated to the virial temperature and fills the halo, so
the accretion rate of the central galaxy in this halo  is a
factor of several below the halo accretion rate.
The filamentary streams in this panel show clumps and
disturbances from substructure. The filaments dragged by the infalling
substructure typically infall with a larger impact parameter bringing
in large amounts of high angular momentum gas.

Results for the $z=1$ halo (right hand panel) are similar, but 
the incoming filaments are somewhat thicker compared to $\rvir$
because a $10^{12}\msun$ halo is no longer an unusually high density
peak.  At fixed overdensity, the physical density and cooling
rates are lower, so a smaller fraction of cold filament gas
survives its journey to the halo center, and more is shock heated to
join the hot atmosphere.  

In the most massive halo in our L12.5/288 simulation, with a mass of
$M_h\approx 9 \times 10^{12}\msun$, the situation is different yet again as we
show in Figure~\ref{fig:mostmassive}.
In this high sigma peak, there are again multiple filaments entering
the virial radius, and the inter-filamentary space is filled with shock
heated gas that extends well beyond $\rvir$. 
Compressed by this hot medium, the filaments are even thinner
compared to the virial radius.
However, the middle and lower panels show that these cold
flows do not penetrate deeper than about $0.5\rvir$.
Some of the filaments are still connected to satellite galaxies, but
the inner satellites and the central galaxy are separated from this gas 
supply.  The bottom, 
zoom-in panel shows that the remnants of the filamentary flows
clump into numerous dense cold clouds. Some of these dense
concentrations of gas are actual galaxies, but most 
are a consequence of fragmentation of the 
filamentary flow induced by 
shocks and high pressure, cooling instabilities of the warm/hot gas, or
clouds that form out of stripped galactic
material. Accretion of such clumps can provide fresh gas for central
galaxies in massive halos, at least in simulations without strong 
preventive feedback. 

\subsection{Accretion rates of individual galaxies}
\label{sec:acc_rates}

\begin{figure*}
\epsfxsize=6.0in
\epsfbox{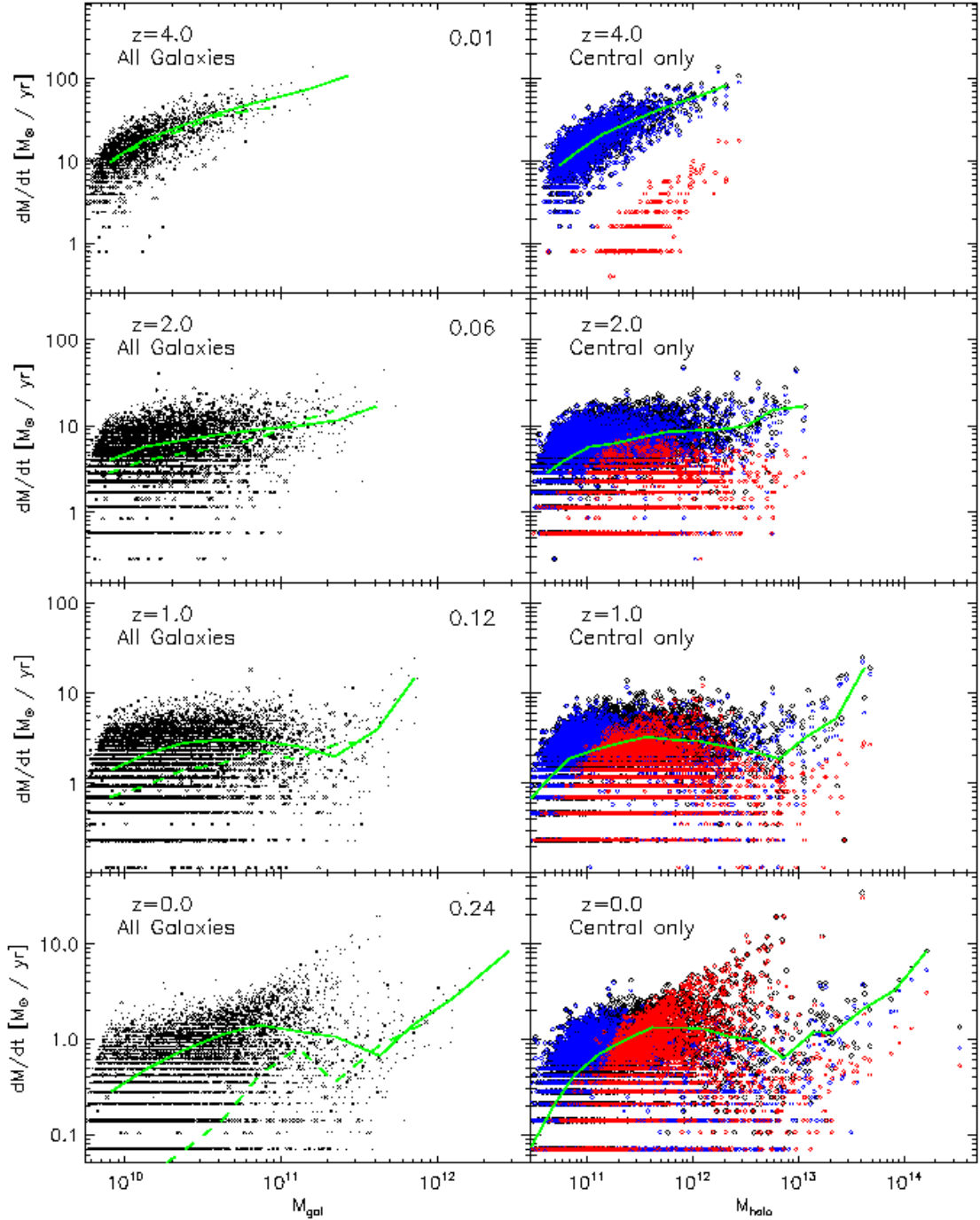}
\caption[Smooth gas accretion rates of individual galaxies in the
  L50/288 simulation]{Smooth gas accretion rates of individual
  galaxies in the L50/288 simulation. In the
  left panels we show the accretion rates of all resolved galaxies as
  a function of their mass and in the right panels we show the accretion
  rates of the central (i.e. the most massive) galaxies in a given
  halo as a function of their parent halo mass. The satellite galaxies
  are marked with crosses, while the central galaxies are plotted as
  points in the left panels and diamonds in the right panels. 
  In both columns, black symbols represent total accretion rates.
  In the right
  panels we also show cold mode accretion in blue and hot
  mode accretion in red. The solid line shows the median accretion
  rate of central galaxies at a given mass while the dashed line is
  for satellites. 
  In the upper right corner of the left panels we also
  indicate the fraction of resolved galaxies with extremely low
  accretion rates
  (see the text for details).
  Note the different vertical scale for the $z=0$ panels.
  }
\label{fig:acc_rates}
\end{figure*}

The accretion rates of individual galaxies as a function of their
mass and of the central galaxies as a function of their parent halo mass
are shown in Figure~\ref{fig:acc_rates}. We also plot the median
accretion rates of the central galaxies as a solid line and the median
accretion rates of satellite galaxies at a given mass as a dashed
line (left panels). 
The larger simulation volume, compared to K05, allows us to highlight 
some details that were
previously not very clear, especially for the high mass end at low
redshifts, which we will discuss shortly. 
 
At $z=4$ all the galaxies are completely cold mode dominated, and the
accretion rates have a steep dependence on galaxy and halo
mass. A typical $10^{10}\msun$ galaxy and a typical central galaxy in a
$10^{11}\msun$ halo accrete at $10-15\msun /\rm yr$, while galaxies in
halos of around $5\times 10^{12} \msun$ accrete at $30-100 \msun / \rm
yr$. \footnote{The typical accretion rates in Figure 7 are lower than
those reported by \cite{dekel09} by a factor of several.
However, they report total gas fluxes through to the inner
15 kpc of their simulated halos rather than the smooth gas
accretion rates of their central galaxies; the difference
between these quantities could be significant.}  
At $z=2$ galaxies at all masses are still cold mode dominated,
but the dependence of the accretion rate on galaxy and halo mass becomes very
shallow.  For a change in halo mass of two orders of magnitude, from
$10^{11}\msun$ to $10^{13}\msun$, the smooth gas accretion rate of 
central galaxies increases by only a factor of $\sim 2$. 
The massive halos at $z=2$ are already full of
hot virialized gas, and only a tiny fraction of this halo gas cools,
resulting in the very low hot mode accretion rates.
Between $z=2$ and $z=1$,
the median accretion rates at the low mass end drop by factors 
of $\sim 2-3$.  Not only are the trends with mass now shallow, but 
above $\sim 2\times 10^{12} \msun$ the accretion rates of the central
galaxies actually start to decrease with mass, though they turn
up again above $10^{13} \msun$. 

In the mass range where the accretion rates are
decreasing with mass, the halos are full of hot virialized gas that
cannot cool.  At the
highest mass end, where the gas accretion rates once again increase with mass,
cold mode accretion dominates. In massive halos only the remnants of the
cold filaments survive at low redshift, in the form of cold
clouds. Some of these clouds are bound and identified with SKID, so
we do not count them as smooth accretion even though they are not
galaxies in the classical sense, i.e. they are not embedded in dark matter
halos. Cold clouds not identified by SKID as galaxies also
contribute to this cold mode accretion in massive halos. 
In addition, some cold, galactic gas gets decoupled from the stellar
component owing to strong ram pressure and tidal forces. 

The accretion of
cold clouds with these various origins causes the cold mode accretion in very
massive halos at late times.
While part of this accretion  could be realistic, much of it 
is probably a consequence of numerical artifacts, for the
reasons discussed in \S\ref{sec:drizzle} below.
Since most
of this accretion comes from clouds with masses slightly below the
resolution limit (see \S\ref{sec:sfr_acc}), we refer to it as
``cold drizzle''. Whether or not this cold drizzle is physical or numerical,
the fact remains that group
and cluster mass halos,  which have masses at least an order of magnitude
above the transition mass between cold and hot mode accretion dominated 
galaxies, effectively stop accreting gas from their hot virialized halos, i.e.
the hot gas no longer cools. 

The results at $z=0$ reveal some interesting trends. 
The dependence of the accretion rates on galaxy and halo mass are
similar to $z=1$, but the rates are lower overall.  At the low mass end
the rates increase from around $0.2-0.3 \msun /\rm yr$ in 
halos with masses of $\sim 10^{11} \msun$ to $1-1.5 \msun /\rm yr$ in
$\sim 10^{12}\msun$ halos. Once again, the
rates drop for masses above $ 10^{12}\msun$. 
At halo masses larger than several times $10^{11} \msun$, the halos 
are completely dominated by hot mode accretion, except
for the most massive halos where potentially spurious cold drizzle provides
several $\msun /\rm yr$ of accreted gas and where the hot mode rates are even
lower. Most halos have lower accretion rates at $z=0$ than at
$z=1$, and smooth accretion is almost completely shut off in halos with
masses larger than several times $10^{12}\msun$. However, there is a
sizeable population of halos with masses of $\sim
10^{12}-5\times10^{13}\msun$ whose 
accretion rates have increased since $z=1$, and 
the mass dependence of the upper envelope of accretion rates
is quite steep. These halos have developed what could be
considered classic cooling flows \citep{fabian94}. 
In the same mass range,  the majority of the
halos have accretion rates, especially in the hot mode, that are very small.
We will discuss the detailed properties and statistics of cooling and 
non-cooling halos in \S\ref{sec:halos}.  

The left panels of Figure~\ref{fig:acc_rates} also list the
fraction of galaxies with extremely low accretion rates, less than 15\%
of the median accretion rate of galaxies at $2\times 10^{10}\msun$. 
Specifically, we adopt thresholds of 
0.1, 0.3, 1 and 3$\msun/{\rm yr}$ at $z=0$, 1, 2, and 4, respectively.
These galaxies typically accrete only one or zero gas particles between the 
two simulation outputs, and most such objects are satellite galaxies in 
larger parent halos.

\subsection{Gas accretion in satellite galaxies}
\label{sec:satellites}

Figure~\ref{fig:acc_rates} includes a comparison of central and
satellite accretion rates in our \newgad, L50/288 simulation.
At $z=4$, the satellites and central galaxies have very similar
accretion rates at a given galaxy  mass. 
At $z=2$, the satellites already accrete
less than the central galaxies, but only a small percentage of galaxies
have their accretion rates drastically reduced. 
The relative difference between central and satellite
galaxies increases  with decreasing redshift and is quite large for
lower mass galaxies at $z=1$. At $z=0$, the differences are substantial. 
The median accretion rate of the lowest mass satellite
galaxies, those around the resolution limit, is zero. Note
that owing to finite particle mass and the time interval between the
two simulated outputs, we cannot measure accretion rates below
0.07$\msun/{\rm yr}$. However, even at
$z=0$ around 45 percent of satellite galaxies are still able to accrete
at rates above 0.1$\msun/{\rm yr}$.
The differences in the  accretion rates of higher mass satellites and
central galaxies of equal mass decrease with galaxy mass, but they are still
noticeable even in satellites with masses of $\sim 2-3 \times 10^{11}\msun$.

At any redshift, roughly 80\% of the galaxies with extremely low
accretion rates are satellites, except at $z=4$ where there is a
roughly equal mix of satellite and central galaxies 
(but at this redshift the low accretion rate galaxies are only 1\%
of the total).
The fraction of resolved galaxies with low accretion rates grows 
from 0.06 at $z=2$ to 0.12 at $z=1$ to 0.24 at $z=0$. 
Although the median accretion rate of all galaxies 
is dropping over this interval, the increase in extremely low accretion
rate galaxies is driven entirely by the growing gap between the
central and satellite populations.
This growing gap is also evident in the median rates --- e.g.,
at $M_{\rm gal} \sim 2 \times 10^{10}\msun$, the ratio of 
median accretion rates is a factor of $\sim 2$ at $z=1$ and
a factor of $\sim 10$ at $z=0$.

\citet{simha08} investigate the relative accretion and merger
rates of satellite and central galaxies in the L22/128, P-TreeSPH
simulation that is the main simulation analyzed by K05, and they
discuss implications for the properties of central and satellite
populations as a function of halo mass.
Although the form of our analysis is different, the general trends
here are consistent with those found in the more detailed investigation
by \citet{simha08}.  In particular, despite the difference
in simulation code, we find that a significant fraction of satellite
galaxies have continuing gas accretion, and (in an investigation not
illustrated here) that the satellites whose accretion is most strongly 
suppressed tend to be low mass systems in high mass halos.
In addition, \citet{mythesis} analyzed satellite accretion in the
cosmological simulations using P-TreeSPH, \newgad and 
Gasoline \citep{wadsley04} SPH codes and also found that the 
fraction of accreting  satellites increases with increasing redshift
and that the fraction of non-accreting  satellites increases with
increasing halo mass. 
These results are consistent with the view that satellites
within the virial boundary of a halo may continue to grow by
accreting gas from massive substructures in which they reside.
These predictions (and the implied continuation of star formation
in satellite galaxies) are supported by the empirical study of
of \citet{weinmann06b}, who find that satellite
galaxies in the SDSS are less red than in semi-analytic models (SAMs) of
galaxy formation.  Especially at high redshifts, our simulation
results indicate that the assumption commonly used in SAMs, that 
satellite galaxies stop accreting fresh gas as soon as they enter a 
larger halo \citep[adopted by, e.g.,][and in subsequent models from these
  groups]{kauffmann93, cole94, somerville99, hatton03}, is not 
valid. At low redshifts this approximation might be correct for the bulk of 
galaxies, but there seems to be a mass dependence on the effect of
satellite strangulation, and a substantial fraction of 
satellites experience continuing accretion.

\subsection{The fraction of accretion in cold and hot mode}

\begin{figure}
\hskip -10pt
\epsfysize=4.9truein
\epsfbox{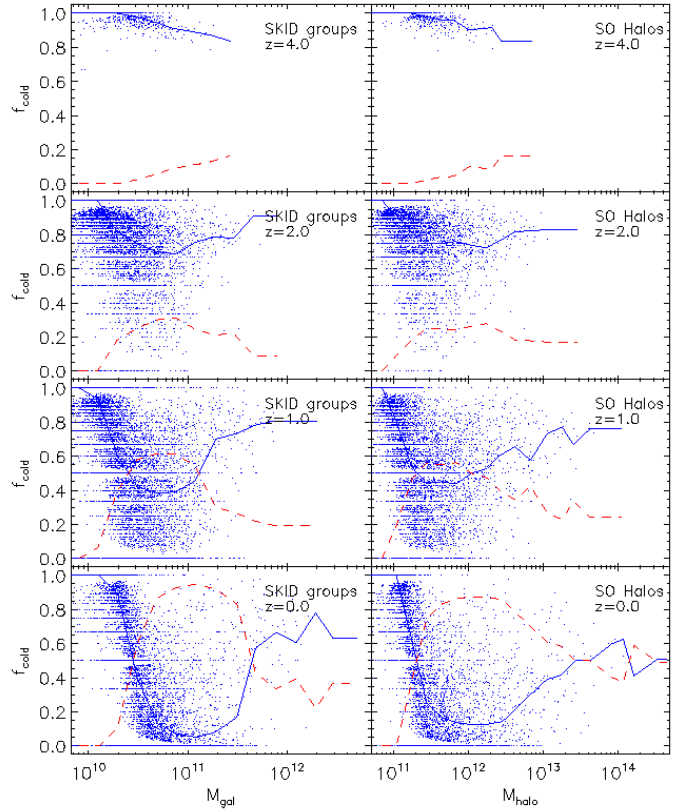}
\vskip -40pt
\caption[The fraction of total smooth gas accretion in cold mode as a
  function of galaxy and halo mass at $z=4$, 2, 1 and 0.]{The fraction
  of total smooth gas  
  accretion in cold mode (blue points) as a function of
  galaxy (left panels) and halo mass (right panels) at $z=4$, 2, 1 and 0.
  The lines represent the median cold mode
  (blue, solid) and hot mode (red, dashed) fractions in a bin 0.2 dex in mass.
}
\label{fig:fractions_j50}
\end{figure}

The relative importance of hot and cold mode
accretion is shown in Figure~\ref{fig:fractions_j50}, where we plot the
fraction of the total smooth gas accretion in these two modes as a
function of galaxy and halo mass.
At $z \ge 2$ the accretion is dominated by the cold
accretion mode at all masses.
At $z=1$, cold mode dominates at the low mass end, i.e in galaxies with 
baryonic masses
lower than $2-3 \times 10^{10}\msun$ or halo masses lower than
$2-3\times 10^{11} \msun$. From 
$M_{\rm halo} \sim 2\times 10^{11}\msun - 10^{12}\msun$,
cold and hot mode accretion are comparable on average.
Above $M_{\rm halo} \sim 2\times 10^{12}\msun$ 
cold mode once again dominates, but the
spread in the cold accretion fraction of individual objects is very large.
At $z=0$, cold mode dominates at low masses, the hot mode
begins to dominate at halo masses $\ge 2-3 \times 10^{11}\msun$, and it
continues
to dominate until masses of $10^{13}\msun$.  Above this mass, the median
cold and hot mode fractions are
comparable, but the spread from halo to halo is very large.
The significant cold mode {\it fraction} at high halo masses arises
from a combination of very low hot mode accretion rates
and ``cold drizzle'' (see \S\ref{sec:acc_rates}), which may
be mostly a numerical artifact.  The absolute rates of cold
accretion in these halos are still fairly small.
We see that when a transition between cold and hot accretion
occurs, the transition masses are
$M_{\rm gal} \sim 2-3 \times 10^{10}\msun$ and
$M_{\rm halo} \sim 2-3 \times 10^{11}\msun$,
very similar to the values in K05.
However, the much lower hot accretion rates in the \newgad
simulation changes the appearance of this plot relative to
K05's figures~5 and~6: cold mode dominates at all galaxy and
halo masses at high redshift, and at low redshift the hot mode
domination at intermediate masses gives way to (possibly spurious)
cold mode domination at the highest masses.

\subsection{Star formation rates}
\label{sec:sfr_acc}

\begin{figure*}
\epsfxsize=6.0in
\epsfbox{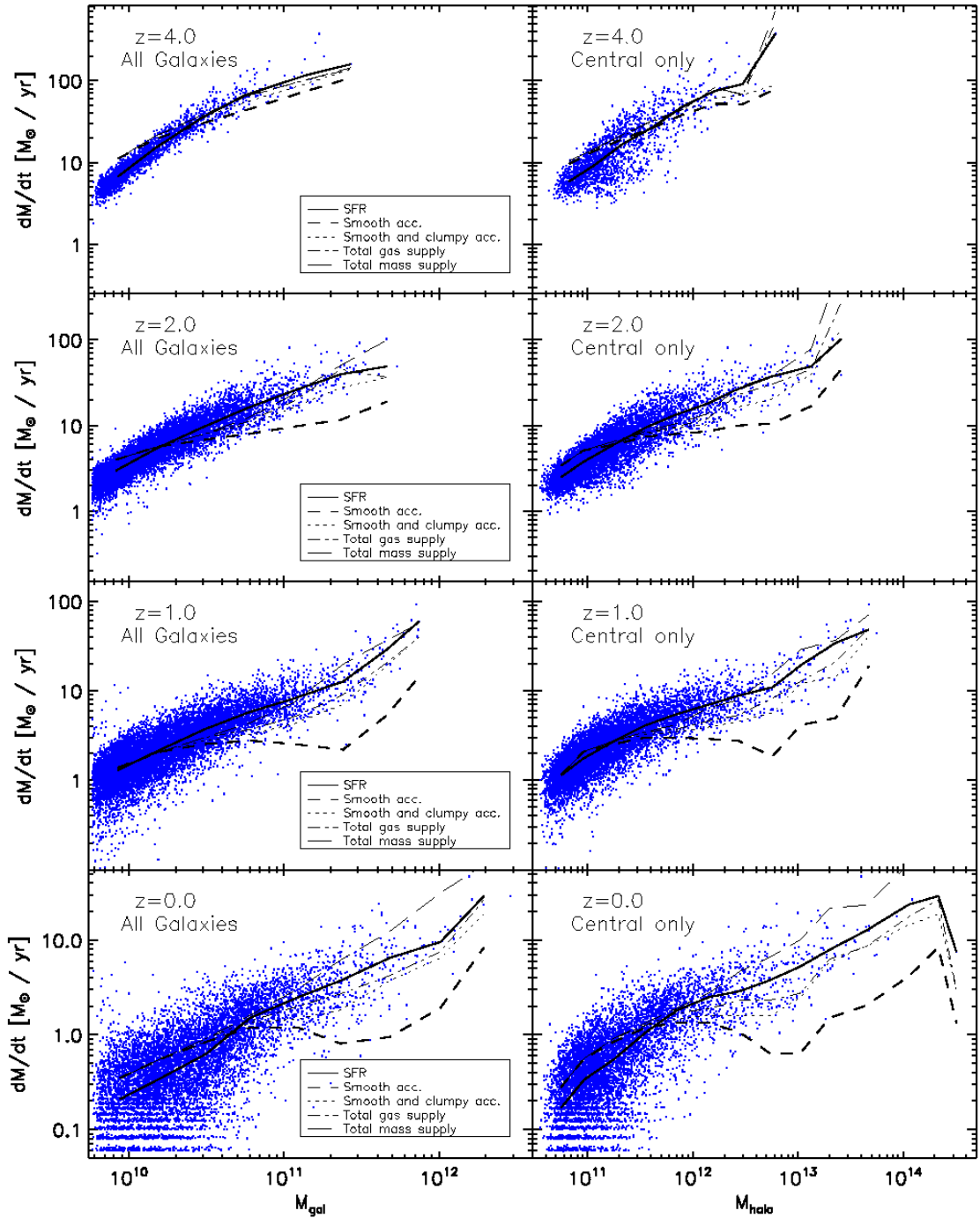}
\caption[]{Star formation rates of individual
  galaxies in the L50/288 simulation. In the
  left panels we show the star formation rates of all resolved galaxies as
  a function of their mass and in the right panels we show the star formation
  rates of the central (i.e. the most massive) galaxies in a given
  halo as a function of the parent halo mass.
  The solid line shows the median star formation rates and the dashed line
  over plots the median smooth gas accretion rates similar to
  Figure~\ref{fig:acc_rates}. The dotted line shows the smooth gas
  accretion rate if we include sub-resolution clumps (see text for
  details). The dot-dashed line includes all channels of gas supply,
  smooth accretion, clumpy accretion and gas delivered through resolved
  mergers, while the long-dashed line also includes the supply of stellar
  component (mostly through mergers).
  The galaxies with zero star formation rates are excluded.
}
\label{fig:sfr_rates}
\end{figure*}

In Figure~\ref{fig:sfr_rates} we show the star formation rates of the
individual galaxies (on the left) and of the central galaxies in halos
(on the right) as a function of the galaxy and halo mass. We plot the median
star formation rate as a solid line and the median smooth accretion
rate from Figure~\ref{fig:acc_rates} as a dashed line. 
We also show the smooth gas accretion rates when we
allow sub-resolution clumps to be counted as accretion (dotted)
and the total gas accretion rates, which includes the gas supplied by
mergers (dot-dashed). In addition we also show median of the total
baryonic mass supply (i.e. including both gas and stars). 
In all four cases we exclude galaxies with zero star formation rate.
The overall star formation rate
follows the smooth gas accretion rate, 
but there are systematic deviations from
this trend. 

In low mass galaxies at low redshift, the
typical gas accretion rates are higher than the star formation rates. 
The accreted gas in low mass galaxies needs
to build up a sufficient reservoir before a significant fraction of the gas
can cross the star forming threshold density.
This physical effect is modulated by a numerical artifact: the gas
densities in the central parts of these marginally resolved objects are
underestimated, which delays the
onset of efficient star formation until even higher masses.
Because of this gas accumulation at the low masses, lower mass galaxies
are always much more gas rich than their higher mass counterparts. As many low
mass galaxies are indeed observed to be more gas rich
at both low \citep{bell00} and high redshift \citep{erb06}, 
part of this effect could indeed be physical, but 
the exact delay in star formation also depends on the resolution. The 
simulations with higher mass resolution show that this behavior shifts
to lower masses as higher densities can be resolved, allowing the star 
formation rate to follow the gas accretion rate more closely. 

Conversely, the star formation rates of high mass galaxies
are typically higher than their accretion rates because of the
larger contribution of minor and major mergers, 
which provide galaxies with gas as well as stars.
In addition, the same objects usually have
slightly higher gas accretion rates at earlier times, 
so the delay between accretion and star formation boosts the latter
relative to the former.
Comparison of the dashed and dotted lines also indicates that a
large amount of the gas accretion in massive galaxies and halos comes from
sub-resolution clumps (cold drizzle), which (as discussed previously)
may be a numerical artifact.
Below $M_{\rm gal} \sim 10^{11}\msun$ or $M_{\rm halo} \sim 10^{12}\msun$, 
adding sub-resolution mergers to smooth accretion makes a
negligible difference. The gas supply from resolved mergers
appears to be only mildly important and only at the high mass
end. Mergers do dominate the total mass buildup of massive
galaxies \citep[e.g.][]{maller06} at late times but most of the mass
they supply is in form of stars.

At all times galaxies form a well defined star forming
sequence with mass, with a slope similar to the observed slope
\citep{daddi07, noeske07}.  However, the normalization of the simulated 
relation is lower than that observed (see \citealt{dave08}).
Similar to the gas accretion rates,
the typical star formation rate for central galaxies in Milky Way
size halos ($\sim 10^{12} \msun$) is $\sim 2\msun/{\rm yr}$
at $z=0$ and grows to $\sim 40 \msun/{\rm yr}$ at $z=4$, albeit with a sizeable
scatter.  The star forming sequence has less 
scatter as a function of galaxy mass than as a function of halo mass. This
indicates that the galactic star formation law also plays a role in the
regulation of this sequence, together with the main driver of gas
supply, cold mode accretion.
As we discuss in Paper II,
reproducing the observed distribution of star
formation rates in high mass galaxies requires that the gas accretion
for most of these systems be sharply reduced but that a fraction
of them retain the accretion and star formation predicted
by the simulation.  

\subsection{Galaxy buildup}

\begin{figure*}
\epsfxsize=5.5in
\epsfbox{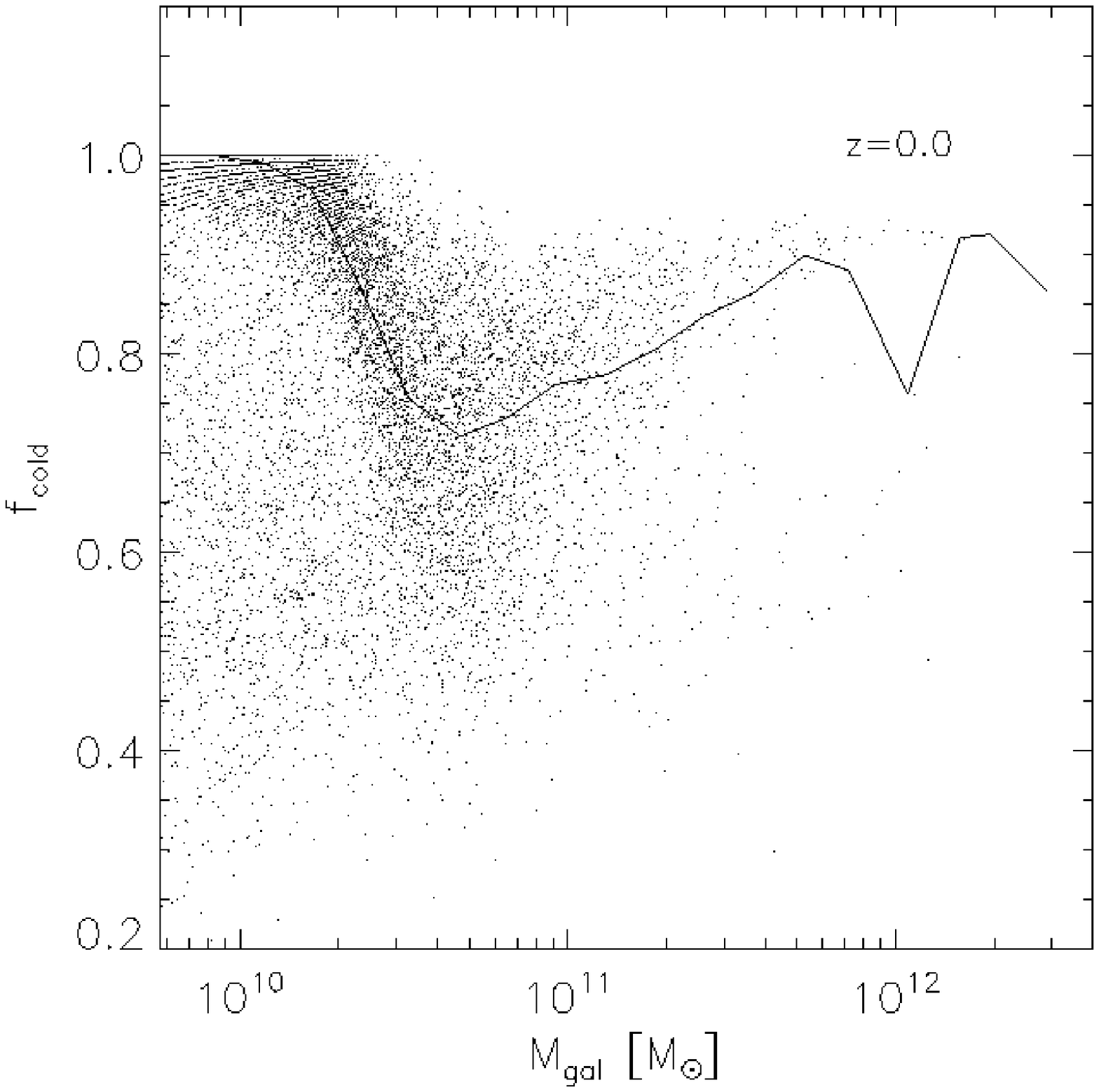}
\caption[The fraction of galactic baryonic mass initially
  accreted through cold mode plotted as a function of galaxy mass
  at $z=0$.]{The fraction of galactic baryonic mass initially
  accreted through cold mode plotted as a function of galaxy mass
  at $z=0$. The solid line plots the median fraction acquired through
  cold mode in bins 0.15 dex in mass.
}
\label{fig:cold_fractions_j50}
\end{figure*}

An intermediate or high mass galaxy may be dominated by hot accretion
at $z=0$, but it spent earlier phases of its growth below the
cold-to-hot transition mass, and at high redshift where all galaxies are 
dominated by cold mode accretion. In addition, a massive galaxy may have 
grown partly by 
mergers with systems that were themselves built by cold mode accretion.
Figure~\ref{fig:cold_fractions_j50} shows the contribution of
gas {\it initially} accreted through cold mode to the {\it final} masses
of galaxies at $z=0$. Here, we select all the particles that are
contained in resolved galaxies  
at $z=0$ and trace their temperature history back through time.
For each particle, we find the maximum temperature that it has ever
reached before that redshift. We exclude any times when the particle's
density was higher than the two-phase medium threshold in \newgad, to
avoid counting temperatures that were increased through the
sub-resolution star formation algorithm. If the temperature of the
particle has never exceeded $250,000\,$K, we count it as being accreted through
cold mode.

As expected, a typical low mass galaxy forms completely through cold
accretion. The contribution of hot mode accretion increases
until $M_{\rm gal} \sim 4-5 \times 10^{10} \msun$, where the cold mode
contribution is $\sim 70-75\%$, but then the trend reverses
and the hot mode contribution starts decreasing in more massive
galaxies. This reversal is caused by the increased contribution of low mass
mergers and mergers at early times, which both add mostly cold mode
accreted material to the massive galaxies.  The near absence of 
hot mode accretion in massive halos reinforces this trend.
The median cold mode contribution at masses 
$M_{\rm gal} \sim 5 \times 10^{11} \msun$ is higher than 90\%,
and it varies between 75\% and 95\% at even higher masses. 
The dispersion in the cold mode fraction is large at all
masses. 
Removing ``cold drizzle'' does not 
alter our conclusion that even the most massive galaxies
are built from baryons that were originally acquired via
cold mode accretion.

The increase of the cold mode contribution at the highest masses
is a consequence of ``downsizing'' in galaxy formation, whereby
progenitors of the most massive galaxies today formed most of
their stars and acquired most of their mass at earlier times than less
massive galaxies.   This trend is established by numerous
observations, and while it is sometimes described as
``anti-hierarchical,''
it is predicted both by semi-analytic models
and by numerical simulations based on the $\Lambda$CDM cosmology.
We will discuss galaxy buildup in more detail in Paper II.
However, it is clear from our results here that
that owing to the high efficiency of
cold mode accretion in lower mass objects, the low efficiency of gas
cooling in massive halos, and the hierarchical buildup
of the high mass objects, the typical galaxy at any mass assembles
primarily through cold mode accretion. While important in some individual
cases and in a limited mass range, cooling from the hot virialized
atmosphere, as emphasized in the standard paradigm \citep{rees77, white78},
represents only a secondary channel of galaxy growth.
While feedback and metal cooling could increase
cooling from the hot atmosphere, as discussed in the next
section, these effects would also increase the mass range over which cold mode
accretion dominates, as discussed in \S\ref{sec:virialization}.
We therefore expect this conclusion to hold even in simulations
that include these processes.

\subsection{Resolution effects}
\label{sec:resolution}

We briefly discuss the effects of resolution on our findings.
For this purpose we use the L11/64 and L11/128 simulations (see
Table~\ref{tbl:sims}), which have the same initial conditions
as the corresponding simulations in K05 but were evolved
using the \newgad code. 
The mass resolution of L11/64 is comparable to our L50/288
simulation, while L11/128 has a factor of 
eight higher mass resolution (a factor of
two in spatial resolution), including additional small scale
power in the initial conditions.  We compare these
simulations only down to $z=1$, since structure on scales larger than
the simulation box becomes nonlinear at lower redshifts. 

Comparing the galaxy
mass function at masses that should be resolved by both simulations,
at the low mass end the L11/128 simulation has a higher mass function while at
the high mass end they both agree quite well. The smooth gas accretion
rates in L11/128 relative to L64
are around 30 percent lower for most halo masses, over the whole
redshift range.
The same applies to the star formation rate. 
These lower rates owe to the higher resolution, which enables lower mass
objects to form at earlier times, converting part of their gas supply into 
stars and hence lowering their gas accretion rates at a given redshift.
With higher resolution the star formation in any galaxy also begins
earlier because higher gas densities can now be reached, and
it therefore locks more baryons into stars at a given time, again decreasing
the amount of gas available for star formation at later times.

Owing to the small volume, there are only two halos more massive than 
$3\times 10^{12} \msun$ at $z=1$ in the L11 simulations.
In the case of central galaxies in these two halos, differences in
the star formation rate are quite small at
$z \ga 2$ but increase to 30-40 percent by $z=1$.
The cooling of hot gas in massive halos is very inefficient in both 
simulations, but it is even less efficient in L11/128,
which is not surprising as a slightly
larger fraction of baryons are now locked up in galaxies.
The effect of ``cold drizzle'' from the sub-resolution
objects, which dominates the supply of gas to massive galaxies at late
times, is similar at high resolution if we count only accretion
below the $64 \times m_{SPH, L11/128}$ threshold, 
but it is significantly
less pronounced in the higher resolution simulation if we count the gas
accretion from groups below $64 \times m_{SPH, L11/64}$.
This result remains the same even if we account for the gas
contribution from resolved mergers (as some of the cold clouds are now
above the resolution threshold), implying that there is a
significant decrease in the total accretion of cold gaseous clouds in
the higher
resolution simulations.  At least part of this effect must owe to the
fact that
galaxies of a given mass are actually more gas poor in the higher
resolution simulations, and there is therefore less gas
available for stripping once they fall into a massive halo, but it
could also suggest that the evolution of the cold clumps is greatly
affected by our limited resolution.
We also find that the typical mass of the cold gas clumps in the halo are
lower in the higher resolution simulation.

The L11/64 vs. L11/128 comparison suggests that our conclusions
about the declining importance of hot mode accretion in massive
halos would if anything be stronger at higher resolution.
They also indicate the robustness of our qualitative conclusions
about the feedback mechanisms needed to reproduce observed
galaxy masses and specific star formation rates, which is the subject
of Paper II.  The significant resolution dependence of the
accretion and star formation rates for the most massive galaxies
implies that the simulation results for these galaxies should
be interpreted with caution.  In particular, these tests
provide some evidence that continuing cold accretion in the
most massive galaxies at late times, is partially a numerical artifact, a
point we return to in \S\ref{sec:drizzle} below.

\section{Cooling and Non-Cooling Halos}
\label{sec:halos}

In this paper, as in K05, we find that galaxy growth in halos
below $M_{\rm halo} \sim 3\times 10^{11} M_\odot$ is dominated
by cold accretion, and that dominant hot gas components develop
only above this mass.  The important difference relative to K05 is
that the hot mode accretion rates in these massive halos are
much lower.  The sign of this difference is as expected given
the difference in SPH algorithms used, with the entropy-conserving
algorithm providing a better separation between the cold and
hot gas phases.  However, as we show below, the cooling rates in 
most of our simulated massive halos are much lower than predicted by
typical semi-analytic calculations in the absence of feedback.
Furthermore, the range of cooling rates is large at fixed halo mass.
In this section we discuss the origin of this large dispersion
of cooling rates, the nature of the continuing cold accretion
in high mass halos, and the possible connections to the observed
properties of X-ray groups and massive galaxies.

\subsection{Gas profiles: cores and cusps}
\label{sec:profiles}

\begin{figure*}
\epsfxsize=6.7in
\epsfbox{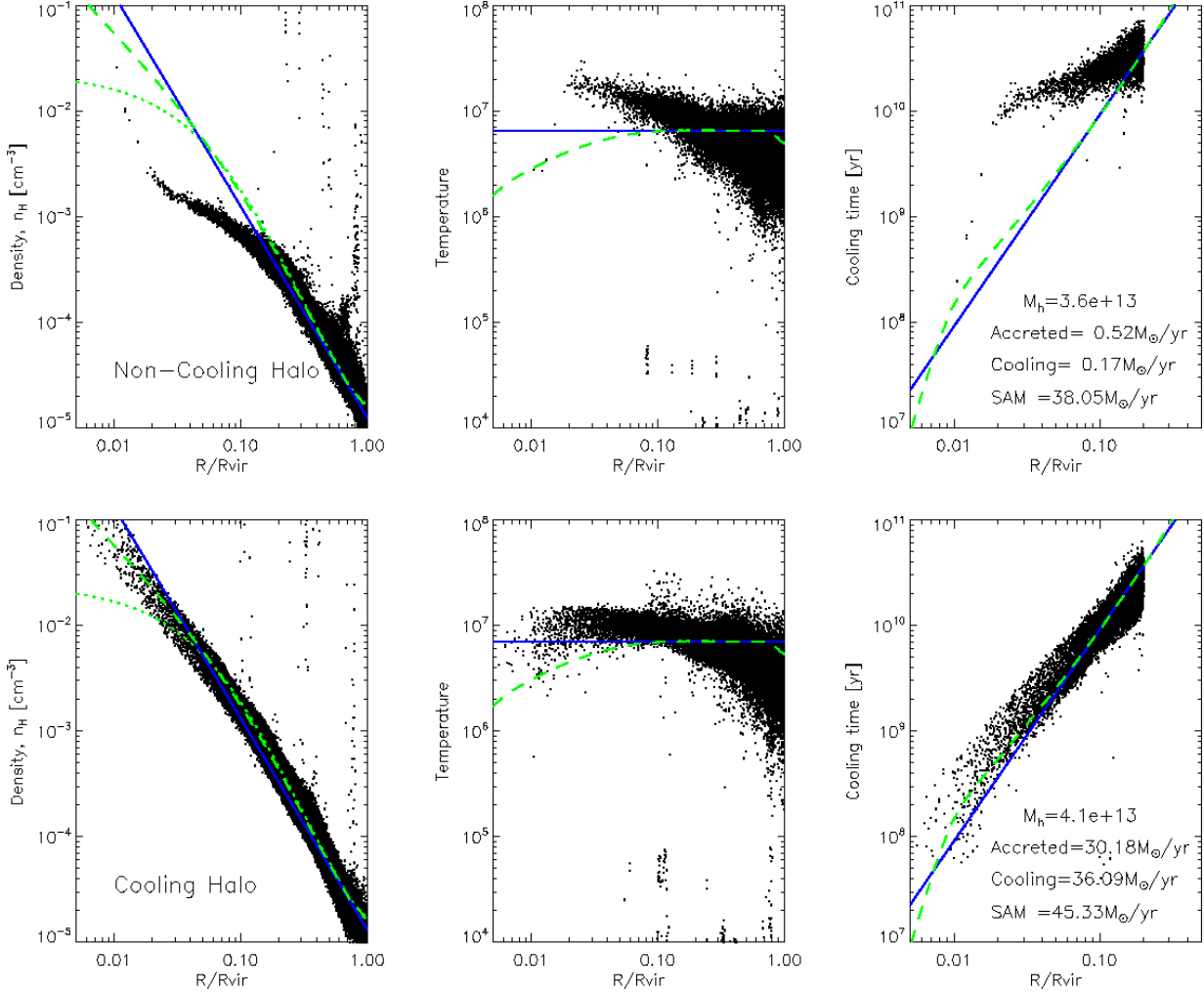}
\caption{The gas properties at $z=0$ of two group size halos with masses of
  $4\times 10^{13} \msun$, one that is not cooling as is typical (upper panels)
  and a rarer one that can cool (lower panels). The left panels
  plot the densities of gas in the halo (points). Model
  profiles are shown for an isothermal gas
  density (solid line) and the density calculated assuming an
  isothermal gas in hydrostatic equilibrium within an  NFW potential for a
  $10^{14}\msun$ halo, where dotted line shows the initial density
  profile and dashed line shows the gas density profile after it cools
  for 2 Gyr. 
  The middle panels plot the gas temperatures and the right panels
  plot the gas cooling times. The right panels also indicate hot mode accretion
  rates from the simulation, an accretion rate from a simple cooling
  calculation, and the cooling rates from the SAM used in Croton et
  al. (2006) (see text for details).
}
          
\label{fig:profiles}
\end{figure*}

The right panels of Figure~\ref{fig:acc_rates} show the accretion
rates of central galaxies, and here we focus on the behavior at
high halo masses.  Above $10^{13} M_\odot$, there are 63 halos
at $z=0$ and 28 halos at $z=1$; at $z=2$ we lower the threshold
slightly to $8\times 10^{12} M_\odot$ to improve statistics,
yielding 3 halos.  At $z \geq 1 $ there is no clear demarcation between 
cooling and non-cooling halos, but if we take an arbitrary division
at $3.5 \msunyr$, then 100, 65, and 13 percent of the massive
halos are above this threshold at $z=2$, 1, and 0, respectively.
If we consider only hot accretion, these percentages change to
33, 0, and 3.

Figure~\ref{fig:profiles} compares the density and temperature
profiles of two massive ($M_{\rm halo} \sim 4\times 10^{13}\msun$)
halos at $z=0$, one in which the central galaxy has a high
hot mode accretion rate of $30\msunyr$ and one with a (more typical)
hot mode accretion rate of $0.3\msunyr$. We only show particles that
are part of a halo but are not part of the individual galaxies
within a halo.  The density profiles
(shown by plotting the densities of individual SPH particles vs.\
radius) are strikingly different: the cooling halo has a roughly
power-law profile in to $0.01 \rvir$, while the non-cooling halo
has a core (more accurately, a sharp break to a shallower power law)
inside $\sim 0.15\rvir$.  The temperature of the hot gas is
roughly constant in to $0.01\rvir$ in the cooling halo,
but in the non-cooling halo it rises by about a factor of two
from $\rvir$ to $0.02\rvir$, inside of which essentially all
gas particles are cold and belong to the central galaxy.
These density and temperature profiles are typical of other high
mass halos in these two regimes.

The right panels of Figure~\ref{fig:profiles} show the instantaneous
cooling times of gas particles in the two halos.
In the cooling halo, particles out to $0.1 \rvir$ have
typical cooling times below $10^{10}$ years.
In the non-cooling halo, only a small fraction of particles have
cooling times below $10^{10}$ years.
We can predict hot mode accretion rates from these plots by 
taking the amount of gas that can cool during the typical
time between simulation outputs ($\sim 1.3$ Gyr at $z \sim 0$) and dividing
by the time interval.  We obtain $36\msunyr$ and $0.17\msunyr$,
in comparison to the measured hot accretion rates of 
$30\msunyr$ and $0.35\msunyr$.  Thus, the difference in hot
mode accretion rates can be understood as a direct consequence
of the difference in density and temperature profiles.
While dynamical processes like shocks,
infall, dynamical friction and turbulent motions likely contributed to
the formation of cored profiles \citep[e.g.][]{conroy08, dekel08,
  khochfar08}, and they may have injected heat that offset
cooling in the earlier stages of halo formation, at late times
the simple ``instantaneous'' estimate of cooling
can explain the very low accretion rates found
in \S\ref{sec:accretion} without the need for any additional 
heating processes.   We briefly checked the higher redshift results
and found that similar cores start to form in the most massive halos
in our simulation even at $z=2$, albeit with slightly higher central
densities. We will explore the profiles of the higher redshift sample
in future work. 

The density
and temperature profiles of these two halos are reminiscent of 
the gas profiles in observed cooling flow and non-cooling flow clusters
\citep[e.g.][]{degrandi02, morandi07}. 
This leaves the question of what creates the range of profiles 
in the first place.
In the cluster context, a common hypothesis is that mergers
disrupt the quiescent gas evolution that would otherwise
produce cooling flows.  Large scale shocks during major mergers
can raise the central entropy of the gas, thereby lowering
the central density and increasing cooling times.
As a rough test of this hypothesis, we compared the mean 
assembly redshifts (defined as the redshift at which the stellar mass
of the largest progenitor exceeded half the mass of the $z=0$ galaxy)
of the central galaxies of massive $z=0$ halos 
with hot mode accretion rates above and below $3.5 \msunyr$. 
These assembly redshifts indicate the time at which the central
parts of the halo were largely in place.
While the scatter is large and the number of systems rather
small, we find that for $M_{\rm halo} > 8 \times 10^{12} M_\odot$ the
mean assembly redshift of cooling halos is $z \approx 1.7$, while
for non-cooling halos it is slightly below $z=1$.
Thus, it appears that halos with a long quiescent phase
after the assembly of their central regions are most able
to build cuspy density profiles that allow substantial
cooling (see also \citealt{burns08}). 
This can be understood from the buildup of the dark matter halos.
The formation of dark matter halos proceeds in two stages, a rapid
growth by major mergers at early 
times and slow growth by minor mergers at late times \citep{li07}. 
The time when the rapid
growth stage finishes increases with increasing halo mass. While
a large fraction of halos in the range of $10^{12}-10^{13} \msun$
typically still experience 
major mergers around $z=1$, most such halos have finished their
rapid growth by $z=0$.  For more massive halos, however, 
a significant fraction
are still experiencing major mergers at $0 \leq z \leq
1$, though others are already in a stage of more quiescent growth.

In this scenario, cooling flows set in when the growth of
halos slows.  Observationally, this view is supported by the
study of \citet{vikhlinin07}, who find that none of the $\sim
20$ clusters they observe in X-rays at $z > 0.5$ show any signature of
strong cooling flow profiles.  At low redshift, however, roughly half
of the clusters in their sample have cooling flow profiles.
In the bulk of clusters in their higher redshift sample, they also 
find some signatures of recent merger activity, and they conclude that 
the higher frequency of major
mergers at higher redshifts is responsible for the difference
in cooling flow fractions.
In our simulation,
we find no high mass halos with hot accretion rates above
$3.5 \msunyr$ at $z=1$.
Cold accretion would not produce X-ray emission, so it is
not relevant to this comparison.
At $z=0$, 3 percent of the $M> 10^{13} M_\odot$ halos have
hot accretion rates above $3.5 \msunyr$, and 17 percent
above $1\msunyr$.  The fraction of cooling flows is thus
low compared to the observations, but the redshift trend
is correct.  
The predictions are likely to depend on halo mass
as well as redshift, so quantitative comparison to data
will require larger volume simulations with more halos
in the mass range ($\sim 10^{14}-10^{15} M_\odot$) probed
by X-ray cluster observations.

Our preliminary tests of simulations with stellar wind feedback and
metal cooling from \citet{oppenheimer06} show that even with the larger
fraction of gas in these halos, gas density cores
are still common in a mass range similar to that in our
simulation. The central parts of these cores are able
to cool at slightly higher rates than in our no-wind, no metal
simulations. Despite the low cooling rates of our present simulations,
additional feedback in group mass halos might be needed to 
keep their massive central galaxies red enough to match the observations.
However, the
formation of a gas density core that prevents or at least
significantly slows down cooling from the hot atmosphere appears to be
a generic result of  hierarchical structure formation, even
without feedback from AGN or other astrophysical sources.
The exact fraction of halos affected, with or
without additional feedback mechanisms, remains an open question.

\subsection{Implications for Semi-Analytic Modeling}
\label{sec:sam}

Returning to Figure~\ref{fig:profiles}, the solid blue lines
show singular isothermal sphere (SIS) gas profiles,
$\rho \propto r^{-2}$, scaled to the virial mass and gas
fraction of these halos.  This density profile describes
the cooling halo fairly well, but it drastically overestimates
the density inside $0.1\rvir$ for the non-cooling halo.  
The solid green curve shows the density profile for isothermal
gas in hydrostatic equilibrium in the profile of an NFW
halo \citep{navarro97} with the concentration expected for
$10^{14} M_\odot$ \citep{bullock01a}, again scaled to the virial
mass and gas fraction of our simulated halos.
We modify the NFW potential to include the gravitational
force softening at small radii, and this modification produces the
flattening of the density profile inside $\sim 0.05\rvir$.  
In Keres et al.\ (2009b, in preparation), we evolve spherically
symmetric gas distributions with these initial profiles
using \newgad.  The dashed green
curve shows the density and temperature profile after the
isothermal NFW distribution has been evolved (with cooling)
for 2 Gyr.  The evolved density profile approximately follows
the SIS curve, while the temperature profile drops inside
$\sim 0.03\rvir$.

The piecewise cooling solution of \citet{bertschinger89}
provides accurate estimates of the gas accretion rates at
halo center for profiles such as these \citep{mythesis}.
Semi-analytic models frequently use an approximation to this
solution, assuming either SIS profiles or isothermal gas
in equilibrium in an NFW halo, to calculate accretion rates
in massive halos.  Comparing the cooling time profiles in
Figure~\ref{fig:profiles}, it is clear that this approach
will be reasonably accurate for the cooling halo but will
overestimate the accretion rate in the non-cooling halo
drastically.  As a specific example, we 
applied the cooling algorithm used in the \cite{croton06}
semi-analytic model: assuming an isothermal gas distribution
and the gas fraction of the simulated halos, we calculate
the accretion rate as $d m_{\rm cool}/ dt= 
0.5 m_{\rm gas}r_{\rm cool}V_c/R_{\rm vir}^2$, 
where $r_{\rm cool}$ is defined as
$t_{\rm cool}(r_{\rm cool})=R_{\rm vir}/V_c$. 
(We ignore the small difference in virial radius definition.)
The predicted accretion rates are $45 \msunyr$ for the cooling
halo, which is accurate at the 50\% level, and $38 \msunyr$
for the non-cooling halo, which is more than two orders of
magnitude above the actual accretion rate in the simulation.
More detailed comparison of the simulation results to a 
full range of semi-analytic models will be presented elsewhere
(Lu et al. 2008).

It is possible that the cores found here in most high mass
halos are enhanced by some numerical artifact caused by the effective
``phase separation'' between hot and cold gas that arises 
in the entropy-conserving formulation of SPH.
However, our tests on idealized cooling flow models
\citep{mythesis} suggest that \newgad gives reasonably
accurate results for halo masses and particle numbers like
those considered here (more accurate than the codes without the entropy
conserving SPH formulation).  Taking the numerical results at
face value, there are significant implications for
semi-analytic models, since with standard assumptions 
they would greatly overestimate the cooling rates for the
majority of high mass halos in this simulation.
Some models reset their cooling radius after a major
merger \citep[e.g.][]{somerville99},  
which goes in the right direction, but an isothermal post-merger 
gas profile still leads to an overestimate of the cooling rates,
by an amount that depends on the time
interval and the algorithm used for the cooling calculation. 
Models that assume that there is a core in the gas density profile of massive
halos \citep[e.g.][]{bower06} could in principle provide a better match to the
central parts of our simulated halos, but the central temperatures in
such  models should also be increased to match the simulated gas properties.
Moreover, models that assume a density core in {\it all} halos fail to 
represent the cooling flow systems that develop at late times.
If semi-analytic models are consistently overestimating
gas cooling rates in massive halos, then they are likely requiring 
too much feedback in these halos to compensate.
More accurate calibration of the cooling recipes in semi-analytic
models can be developed by comparing the evolution of gas
profiles in hydrodynamic simulations and the semi-analytic calculations;
the short analysis here provides a first step in this direction.

\subsection{Cold Drizzle}
\label{sec:drizzle}

Even at high halo masses and low redshifts, accretion onto 
central galaxies is usually dominated by cold gas in 
sub-resolution clumps, the form of accretion that we have
described as ``cold drizzle.''  
Thermal instability can form cold clumps 
{\it in situ} from hot halo gas \citep{mo96,maller04}, 
but the cold drizzle gas in our simulations has
{\it never} been hot, so thermal instability clouds
are not its source.  (Our simulation probably lacks
the resolution needed to model thermal instability
in galaxy mass halos in any case.)
Rather, the cold drizzle appears to be a mix of 
gas that has been stripped from larger galaxies and sub-resolution
clumps that form within cold/warm filaments that are disrupted at the
outskirts of high mass halos, where they may be compressed by the
surrounding hot gas. 

Both of these sources should produce cold gas that can accrete
onto galaxies in massive halos.
However, there are several numerical effects that could artificially
amplify this form of accretion in our simulation, perhaps
by large factors.
First, ram pressure stripping of cold gas from infalling galaxies
\citep{gunn72} is almost certainly overestimated at
the resolution of L50/288 because the typical smoothing 
lengths of the hot particles greatly exceed the size of the galaxies
themselves (see tests and discussion by \citealt{titley01}).
Second, a similar effect could cause cold gas clumps (regardless
of origin) to sink to the center of the halo more quickly than
they should, as artificial ram pressure slows their tangential
velocities.  Finally, SPH codes have difficulty 
following surface instabilities that should develop 
when cold dense clumps move through a hot dilute medium \citep{agertz07},
so clouds that should be destroyed by instabilities or by conduction
(which is not included in our simulations)
may artificially survive their journeys to the halo center.
As discussed in \S\ref{sec:resolution}, we find that the
amount of cold drizzle in massive halos at $z=1$ is reduced
when the mass resolution is increased by a factor of eight
relative to that of L50/288, though the limited redshift range
and box size of our resolution tests does not allow us to draw
clear conclusions about the influence of resolution on 
cold accretion rates in massive halos.
In a systematic comparison of simulations run with different
codes from the same initial conditions \citep{mythesis}, we find more
cold drizzle in massive halos in simulations run with \newgad
compared to those run with PTreeSPH and Gasoline.
This difference suggests that the effects mentioned above
may be more severe for the \newgad SPH formulation because of 
the different calculation of the hydrodynamic forces and the different
averaging of the properties of gas on the interface between the dense
cold phase and the hot dilute medium, which could result in
different rates of cloud survival and formation and different ram
pressure forces at moderate resolution. 

Our findings on low accretion rates in massive halos are qualitatively
similar to those of \citet{naab07}, based 
on much higher resolution \newgad simulations of the formation 
of individual galaxies in a cosmological context.  
Like the simulation analyzed here, the \citet{naab07} 
simulation incorporates a simple star formation algorithm
and has no strong feedback in the form of supernova-driven winds
or AGN heating.  At their highest resolution, \citeauthor{naab07}
find that the central galaxy in a halo of several 
times $10^{12} \msun$ stops forming stars and is not supplied with
new gas after a redshift $z \approx 1$; thus they, too, find
that accretion from the hot gas halo shuts off without the
assistance of feedback.
In their lower resolution simulations, however, accretion
continues at lower redshift, perhaps in the form of cold drizzle
like that in our simulations.  Increased resolution could suppress
this cold accretion both by more accurately treating ram pressure
and instabilities and by depleting the gas supply in low mass
clumps (which get converted into stars when the resolution is
high enough).

We remain uncertain whether the cold accretion rates in massive halos
in our simulation are approximately correct or substantially
overestimated because of finite resolution and numerical effects
in the \newgad SPH formulation.  
We show in Paper II that cold drizzle makes only 
a small contribution to the final masses of galaxies in
these high mass halos, but it can 
significantly affect their star formation rates and
hence their colors.  

\subsection{A Natural End to Smooth Accretion?}
\label{sec:natural}

Following the results of \cite{katz03} and \cite{birnboim03}
on cold and hot accretion,
several authors noted that simulations or semi-analytic models
of galaxy formation would better reproduce observations of 
the galaxy luminosity function and the low star formation rates
of massive galaxies if hot accretion were suppressed
(\citealt{binney04}; K05; \citealt{dekel06}; \citealt{cattaneo07}).
Motivated in part by observations of galaxy clusters,
these authors and others suggested that preventive feedback from AGN 
might preferentially suppress hot accretion, since relative
to cold accretion the hot gas has lower density, higher temperature,
and larger geometrical cross-section.
In one concrete implementation of this idea, 
\cite{cattaneo06} and \cite{cattaneo08}
show that a semi-analytic model that assumes the
shut-down of accretion and star formation in halos above
a critical mass $M_{\rm halo} \sim 10^{12} M_\odot$, 
together with traditional semi-analytic model assumptions
about stellar feedback, reproduces many of the observed
features of the galaxy red sequence and bimodality of
the galaxy population.  The semi-analytic models of
\cite{croton06} and \cite{bower06} allow AGN feedback
to suppress gas accretion in halos that, according to the
cooling radius criteria in these models, 
have developed quasi-static hot gas halos \citep{white91}, and these
models also achieve better agreement with observations
than earlier generations of semi-analytic models.

In this paper we find that the hot accretion rates in
simulated massive halos are low even without any 
strong feedback mechanisms, principally because of the
formation of cored gas density profiles that 
have long central cooling times.  If we assume that
cold drizzle in massive halos is mostly a numerical
artifact that would disappear at high resolution
(\S\ref{sec:drizzle}), we conclude that the physics
of hierarchical halo assembly and galaxy formation 
on its own leads to a shutdown of gas accretion in 
massive halos.  \cite{naab07} emphasize that the simulated
galaxies that form in their highest resolution 
simulations have many of the properties of observed massive
ellipticals --- in particular, no ongoing gas accretion
and star formation.  Maybe smooth accretion comes to
a natural end, with no need for AGN feedback or other
suppression mechanisms.

However, the story does not appear to be quite so 
simple.  In our simulations, the fraction of cooling
halos increases towards lower masses, and while we
find some density cores even in halos of $\sim 10^{12} \msun$,
core formation is not enough to shut down gas cooling in
most halos in the $10^{12}\msun - 10^{13}\msun$ range.
If we remove hot mode accretion and cold drizzle in massive halos ``by hand,''
then we find that the most massive galaxies are predominantly
``red and dead,'' in agreement with the high resolution
results of \cite{naab07}.  However, in Paper II we show that
the simulated massive galaxies are too massive by a factor
of $2-5$ compared to observed galaxies of the same space density,
so in the absence of feedback the processes simulated here
and by \cite{naab07} do not appear able to reproduce the
galaxy baryonic mass function; at a given halo mass, galaxies
in the real universe are less massive.
On the other hand,
we also show in Paper II that reproducing observed specific
star formation rates requires that central galaxies in a
sizeable fraction of $10^{12}-10^{13}\msun$ halos remain on
the star forming sequence, so a complete shutdown of accretion
above $10^{12}\msun$ is not what is desired.
Finally, we note again that ejective, galactic wind feedback
in lower mass galaxies will ultimately have significant
impact on the most massive galaxies by reducing the stellar
and gas masses of galaxies that merge with them, by
increasing the overall IGM gas supply, and by enriching
the IGM and therefore enhancing cooling rates.
Our findings here are important steps towards understanding
the physics of galaxy growth, and they differ in some significant
ways from conventional wisdom, but they do not provide
a full solution to the puzzles of observed galaxy evolution.

\section{Conclusions}
\label{sec:conclusions}

Our principal results are based on a cosmological SPH simulation
of a $50\hmpc$ (comoving) box in a $\Lambda$CDM universe,
evolved with $288^3$ dark matter and $288^3$ gas particles
using the code \newgad.  The combination of volume and particle
number allows us to resolve (with $> 64$ particles) galaxies of
baryonic mass $\mgal \ga 6 \times 10^9\msun$ while following
the formation of several cluster-size halos with $\mhalo \ga 10^{14}\msun$.
Relative to the main simulation analyzed by K05, our present simulation has
an eleven times larger volume, and it uses the entropy-conserving
SPH formulation of \cite{springel02}, which should provide a 
more accurate treatment of multi-phase gas than the 
PTreeSPH code employed by K05.  We incorporate photoionization
by the UV background and thermal feedback on a 2-phase galaxy
interstellar medium, but we do not explicitly add galactic
winds or track metal enrichment, so this is a ``minimal
feedback'' simulation.

Confirming the key result of K05, we find that galaxies with
baryonic masses below $2-3 \times 10^{10}\msun$ or halo masses
below $2-3 \times 10^{11}\msun$ gain most of their mass
through ``cold'' accretion of gas that has never been
close to the halo virial temperature
(see also \citealt{kay00,fardal01,katz03}).
In halos below $\sim 2.5 \times 10^{11}\msun$, the
majority of intergalactic gas is colder than 250,000$\,$K,
while in higher mass halos the majority of gas is close 
to the virial temperature; the transition mass stays nearly
constant from $z=3$ to $z=0$. Similar behavior is found in
spherically symmetric calculations \citep{birnboim03,dekel06},
and the likely cause of the transition is that the short
cooling times in low mass halos prevent the formation of
a stable virial shock (\citealt{birnboim03}; see also
\citealt{binney77,white91}).  However, as shown by K05
and illustrated more comprehensively here, halos near and
above the transition mass have hot gas halos penetrated
by cold gas filaments, which may continue to feed the
central galaxy.  Similar results are found in adaptive mesh
refinement simulations (\citealt{ocvirk08,dekel09};
see also \cite{dekel06} figure 6).  
At high redshift these cold filaments penetrate close to 
the halo center, while in high mass halos at $z<1$ they
are usually truncated or disrupted before reaching the center.
The difference likely reflects the higher physical densities
and shorter associated cooling times of the high redshift
filaments that prevent the expansion of a stable shock
\citep{birnboim03} in these high density regions.

The key difference between our present results and those of K05 is 
that {\it hot} accretion rates are much lower, a consequence
of changing to the more accurate, entropy-conserving SPH
formulation.  Because of the low hot accretion rates, the growth
of high redshift galaxies ($z \geq 2$) is dominated by
cold accretion at all galaxy and halo masses.
At lower redshifts, massive halos experience a combination
of hot accretion and ``cold drizzle,'' accreting lumps of
cold gas below our 64-particle galaxy resolution threshold.
Our resolution tests here and those of \cite{naab07}
suggest that some or even most of this ``cold drizzle''
may be a numerical artifact.  Eliminating it would significantly
affect the predicted star formation rates and colors of
massive galaxies, but it would not have much impact on
their predicted stellar masses (Paper II).
At {\it all} galaxy masses, the majority of galaxy
baryonic mass in our simulation was originally acquired
by cold accretion, either directly onto the main progenitor
or onto a satellite that merged with it.

Galaxy accretion rates decline systematically from 
$z=4$ to $z=2$ to $z=1$ to $z=0$.  For 
$10^{10} \msun < \mgal < 10^{11}\msun$, there is a
trend of increasing accretion rate with increasing
mass at all redshifts, though it becomes flatter
at low $z$.  At higher masses, accretion rates have large
scatter and are sensitive to cold drizzle.
With our adopted star formation prescription,
empirically motivated by the Kennicutt-Schmidt law
\citep{kennicutt98,schmidt59}, star formation closely
tracks gas accretion with a short delay; K05 showed this
on a global basis, and here we show it on a galaxy-by-galaxy
basis.  At each redshift, more massive galaxies tend to
have higher star formation rates but (moderately) lower
specific star formation rates.  Satellite galaxies
orbiting within the virial radius of larger halos
experience continuing accretion and star formation.
The differences in the accretion rates of central and satellite
galaxies of the same mass are small at high redshift,
while at low redshift a significant fraction of satellites
have much lower accretion rates.  More detailed analysis
of the satellite accretion and merger rates (based on
the K05 simulation) appears in \citet{simha08}.

The low hot accretion rates in our simulated massive halos
arise because these halos typically have cores in their hot
gas density profiles (or, more precisely, because they
break from an approximately $r^{-2}$ profile to a much
shallower profile in their inner regions).  By $z=0$,
a modest fraction of the $M>10^{14} M_\odot$ halos develop
cuspy density profiles with cooling cores, evolution that
is reminiscent of Vikhlinin et al.'s (\citeyear{vikhlinin07})
finding that cooling flow clusters are fairly common at $z=0$
but much rarer at $z>0.5$.  We speculate that the cores are
produced by shocks during the chaotic, rapid assembly phase
of halos, and that the more quiescent evolution during the
slow accretion phase allows cooling to establish a
cusped profile.  
The accretion rates of central galaxies in our massive
halos are predicted with reasonable accuracy by
instantaneous cooling calculations that use the actual
simulated gas density profiles, but cooling calculations
that assume isothermal gas in equilibrium in an NFW halo or
singular isothermal halo may overestimate the cooling rates
by orders of magnitude.  Our simulation predicts low
hot accretion rates even though it does not include
AGN heating or other forms of preventive feedback.
While such feedback may be needed to explain the absence
of cold gas in cooling flow clusters
\citep{kaastra01,peterson03}, semi-analytic models may
overestimate its global importance by assuming incorrect
halo gas profiles (and thus excessive cooling).

The comparison to observations in Paper II shows that this simulation,
despite its low hot accretion rates, overpredicts the galaxy
baryonic mass function.  The amount of baryons that is locked in
  galaxies at $z=0$ and the amount of baryons locked in stars at $z=0$
  is about a factor of 2-3 higher than observed. This is caused by simulated
  galaxies being too massive at all masses, but the exact difference
  is a function of galaxy mass with the lowest and the highest mass
  galaxies being the most problematic. 
Matching observations therefore
requires either the suppression of cold accretion (which seems
unlikely on physical grounds) or substantial amounts of 
ejective feedback that returns gas to the IGM before it forms stars.
Such ejective feedback is also needed to explain the observed
enrichment of the IGM (e.g., \citealt{oppenheimer06,oppenheimer08}).
Ejected, metal-enriched gas will affect the cooling of the
IGM in massive halos, and it will reduce the rate at which mergers
add stars and gas to massive galaxies.  We will investigate the 
impact of winds and metal-enrichment on galaxy growth
in future work with simulations that
include these effects.  However, our present results suggest that
the cooling of shock-heated, virialized gas, which has been the
focus of many analytic models of galaxy growth spanning more than
three decades, might be a relatively minor element of
galaxy formation.

We thank V. Springel for making the \newgad public and
K. Finlator and B. Oppenheimer for help during the modification of the
public version of the code.
D.K acknowledges the support from the ITC Fellowship at the Institute
for Theory and Computation at the Harvard College Observatory.
We also acknowledge support from NSF grant AST-0205969 and from NASA grants
NAGS-13308 and NNG04GK68G.

\bibliographystyle{mn2e}
\bibliography{}

\end{document}